\documentclass[twocolumn,showpacs,aps,prd,nofootinbib]{revtex4}
\usepackage{graphicx}
\usepackage{dcolumn}
\usepackage{amsmath}
\usepackage{epsfig}
\usepackage{colordvi}
\usepackage{color}
\usepackage{hhline}

\RequirePackage{xspace}

\newcommand{\BABARPubYear}    {12}
\newcommand{\BABARPubNumber}  {030}
\newcommand{\SLACPubNumber} {15324}

\usepackage{relsize}
\def\babar{\mbox{\slshape B\kern-0.1em{\smaller A}\kern-0.1em
    B\kern-0.1em{\smaller A\kern-0.2em R}}}
     
\mathchardef\Upsilon="7107
\def\Y#1S{\ensuremath{\Upsilon{(#1S)}}\xspace}

\def\pep2{PEP-II}

\long\def\inst#1{\par\nobreak\kern 4pt\nobreak
  {\it #1}\par\vskip 10pt plus 3pt minus 3pt}
  
\begin{document}

\begin{flushleft}
SLAC-PUB-\SLACPubNumber \\
\babar-PUB-\BABARPubYear/\BABARPubNumber \\
\end{flushleft}                                                                 

\title{\large \bf
\boldmath
Study of $e^+e^-\to p\bar{p}$ via initial-state radiation at \babar\
}
%% author list as of 07-Oct-2012 (349 authors)
%
\author{J.~P.~Lees}
\author{V.~Poireau}
\author{V.~Tisserand}
\affiliation{Laboratoire d'Annecy-le-Vieux de Physique des Particules (LAPP), Universit\'e de Savoie, CNRS/IN2P3,  F-74941 Annecy-Le-Vieux, France}
\author{E.~Grauges}
\affiliation{Universitat de Barcelona, Facultat de Fisica, Departament ECM, E-08028 Barcelona, Spain }
\author{A.~Palano$^{ab}$ }
\affiliation{INFN Sezione di Bari$^{a}$; Dipartimento di Fisica, Universit\`a di Bari$^{b}$, I-70126 Bari, Italy }
\author{G.~Eigen}
\author{B.~Stugu}
\affiliation{University of Bergen, Institute of Physics, N-5007 Bergen, Norway }
\author{D.~N.~Brown}
\author{L.~T.~Kerth}
\author{Yu.~G.~Kolomensky}
\author{G.~Lynch}
\affiliation{Lawrence Berkeley National Laboratory and University of California, Berkeley, California 94720, USA }
% ...
\collaboration{The \babar\ Collaboration}
\noaffiliation

\begin{abstract}
The process $e^+e^-\to p\bar{p}\gamma$ is studied using 469~fb$^{-1}$ of
integrated luminosity collected with the \babar\ detector at the PEP-II 
collider, at an $e^+e^-$ center-of-mass energy of 10.6~GeV.
From the analysis of the $p\bar{p}$ invariant mass spectrum, the energy 
dependence of the cross section for  $e^+e^-\to p\bar{p}$ is measured from 
threshold to 4.5~GeV. The energy dependence of the ratio of electric and 
magnetic form factors, $|G_E/G_M|$, and the asymmetry in the proton angular 
distribution are measured for $p\bar{p}$ masses below 3~GeV. We also measure
the branching fractions for the decays $J/\psi \to p\bar{p}$ and 
$\psi(2S) \to p\bar{p}$.
\end{abstract}


\maketitle

\setcounter{footnote}{0}

\section{ \boldmath Introduction\label{intro}}

\begin{figure}
\includegraphics[width=.4\textwidth]{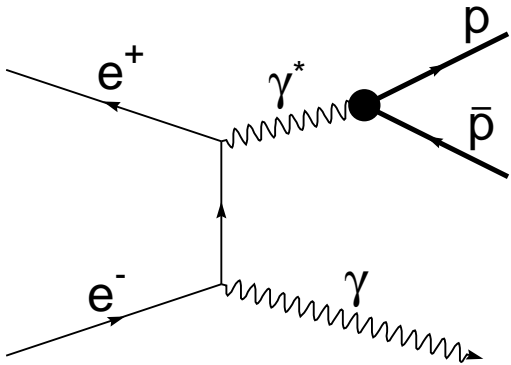}
\caption{The Feynman diagram for the signal ISR process $e^+e^-\to p\bar{p}\gamma$.
\label{fig1}}
\end{figure}

In this paper we use the initial-state-radiation (ISR) technique
to study the $e^+e^-\to p\bar{p}$ process  
in a wide range of center-of-mass (c.m.) energies.
The Born cross section for the ISR process 
$e^+e^-\to p\bar{p}\gamma$ (Fig.~\ref{fig1}),
integrated over the nucleon momenta, is given by
\begin{equation}
\frac{{d}^2\sigma_{e^+e^-\to p\bar{p}\gamma}(M_{p\bar{p}})}
{dM_{p\bar{p}}\,d\cos{\theta_\gamma^\ast}} =
\frac{2M_{p\bar{p}}}{s}\, W(s,x,\theta_\gamma^\ast)\,\sigma_{p\bar{p}}(M_{p\bar{p}}),
\label{eq1}
\end{equation}
where $\sigma_{p\bar{p}}(m)$ is the Born cross section for the nonradiative  
process $e^+e^-\to p\bar{p}$,
$M_{p\bar{p}}$ is the $p\bar{p}$ invariant mass, $\sqrt{s}$ is the nominal $e^+e^-$ 
c.m.~energy,
$x\equiv{2E_{\gamma}^\ast}/\sqrt{s}=1-{M_{p\bar{p}}^2}/{s}$, 
and $E_{\gamma}^\ast$ and $\theta_\gamma^\ast$
are the ISR photon energy and polar angle, respectively,
in the $e^+e^-$ c.m.~frame.\footnote{Throughout this paper,
the asterisk denotes  quantities in the $e^+e^-$ center-of-mass frame.
All other variables except $\theta_p$ are defined in the
laboratory frame.}
The function~\cite{BM} 
\begin{equation}
W(s,x,\theta_{\gamma}^\ast)=
\frac{\alpha}{\pi x}\left(\frac{2-2x+x^2}{\sin^2\theta_{\gamma}^\ast}-
\frac{x^2}{2}\right)
\label{eq2}
\end{equation}
is the probability of ISR photon emission for 
$\theta_{\gamma}^\ast\gg m_e/\sqrt{s}$, where
$\alpha$ is the fine-structure constant and $m_e$ is the electron mass.
The cross section for the $e^+e^-\to p\bar{p}$ process is given by
\begin{equation}
\sigma_{p\bar{p}}(M_{p\bar{p}}) = \frac{4\pi\alpha^{2}\beta C}{3M_{p\bar{p}}^2}
\left [|G_M(M_{p\bar{p}})|^{2} + \frac{2m_p^2}{M_{p\bar{p}}^2}|G_E(M_{p\bar{p}})|^{2}\right],
\label{eq3}
\end{equation}
where $m_p$ is the nominal proton mass, $\beta = \sqrt{1-4m_p^2/M_{p\bar{p}}^2}$, and $C$
is the Coulomb correction factor (see, for example, Ref.~\cite{Coulomb} and references therein), which 
makes the cross section nonzero at threshold
($C = y/(1-e^{-y})$ with $ y = \pi\alpha/\beta$). 
The cross section depends on the magnetic form factor ($G_M$) and
the electric form factor ($G_E$); at threshold, $|G_{E}| = |G_{M}|$.
From the measurement of the cross section a linear
combination of the squared form factors can be determined. 
We define the effective form factor 
\begin{equation}
|F_p(M_{p\bar{p}})|=\sqrt{\frac{|G_M(M_{p\bar{p}})|^2 + {2m_p^2}/{M_{p\bar{p}}^2}|G_E(M_{p\bar{p}})|^2}
{1+{2m_p^2}/{M_{p\bar{p}}^2}}},
\label{eq4}
\end{equation}
which is proportional to the square root of the measured 
$e^+e^-\to p\bar{p}$ cross section.

The modulus of the ratio of the electric and magnetic form factors
can be determined from  the distribution of $\theta_p$,
the angle between the proton momentum in the $p\bar{p}$
rest frame and the momentum of the $p\bar{p}$ system in the $e^+e^-$ 
c.m.~frame. This distribution can
be expressed as a sum of terms proportional to
$|G_M|^2$ and $|G_E|^2$. The full differential cross section for
$e^+e^-\to p\bar{p}\gamma$  can be found, for example,
in Ref.~\cite{kuhn_pp}. The $\theta_p$ dependences of
the $G_E$ and $G_M$ terms are close to 
$\sin^{2}\theta_p$ and $1+\cos^{2}\theta_p$,
respectively.

   Direct measurements of the $e^+e^-\to p\bar{p}$ cross section
have been performed in $e^+e^-$
experiments~\cite{DM1,DM2,ADONE73,FENICE,BES,CLEO,CLEO2012}. 
Most of these measurements have an accuracy of (20--30)\%.
The cross section and the proton form factor were deduced
assuming  $|G_E|=|G_M|$, and the measured proton
angular distributions \cite{DM2,FENICE,CLEO2012} did not contradict
this assumption.
More precise measurements of the proton form factor 
have been performed in
$p\bar{p}\to e^+e^-$ experiments~\cite{LEAR,E760,E835}.
In the PS170 experiment~\cite{LEAR} at LEAR,
the proton form factor was measured from threshold
($p\bar{p}$ annihilation at rest) up to a c.m.~energy 2.05~GeV. The
ratio $|G_E/G_M|$ was measured with about 30\% accuracy
and was found to be compatible with unity.
The LEAR data show a strong dependence of the form factor
on c.m.~energy near threshold, and very little dependence
in the range 1.95--2.05~GeV. Analyses from
Fermilab experiments E760~\cite{E760}
and E835~\cite{E835} show a strong decrease of the form factor
at c.m.~energies higher than 3 GeV, in agreement with perturbative QCD, which predicts
that the  dependence should be  $\alpha_s^2(m^2)/m^4$.
However, the recent precision $e^+e^-$ measurement~\cite{CLEO2012} based
on CLEO data
indicates that the decrease of the form factor at energies 
about 4~GeV is somewhat slower.

The previous \babar\ study~\cite{babarpp} of the process
$e^+e^-\to p\bar{p}$
using the ISR technique was based on about half of the data that 
were finally collected in the
experiment. The $e^+e^- \to p\bar{p}$ cross section
was measured for c.m.~energies up to 4.5~GeV.
This measurement yielded a significant improvement in precision
for energies below 3~GeV.
 In contrast to previous $e^+e^-$ and $p\bar{p}$
experiments, the \babar\ measurement did not assume that 
$|G_E|=|G_M|$. The ISR approach provides Á full $\theta_p$ coverage, and hence
high sensitivity to $|G_E/G_M|$. 
The energy dependence of the form-factor ratio $|G_E/G_M|$ was
measured for c.m.~energies below 3~GeV.
For energies up to 2.1~GeV, this ratio was found to be
significantly greater than unity, in disagreement with the PS170 
measurement~\cite{LEAR}.

In this work we update the analysis of Ref.~\cite{babarpp} using
the full \babar\ data sample collected at and near the $\Upsilon$(4S)
resonance. 

\section{ \boldmath The \babar\ detector and data samples}
\label{detector}
We analyse a data sample corresponding to
an integrated luminosity of 469~fb$^{-1}$ recorded with
the  \babar\ detector~\cite{ref:babar-nim} at the SLAC \pep2\ 
asymmetric-energy collider. At \pep2, 9-GeV electrons collide with 
3.1-GeV positrons at a c.m.~energy of 10.58~GeV 
(the $\Upsilon$(4S) mass). About 10\% of the data are collected
at 10.54 GeV. 

Charged-particle tracking is
provided by a five-layer silicon vertex tracker (SVT) and
a 40-layer drift chamber (DCH), operating in a 1.5-T axial
magnetic field. The transverse momentum resolution
is 0.47\% at 1~GeV/$c$. The position and energy of a photon-produced
cluster are measured with a CsI(Tl) electromagnetic calorimeter, 
which yields an energy resolution of 3\% at 1~GeV. Charged-particle
identification is provided by specific ionization ($dE/dx$) 
measurements in the SVT and DCH, and by an internally reflecting 
ring-imaging Cherenkov detector (DIRC). Muons are identified in
the solenoid's instrumented flux return (IFR), which consists of iron plates 
interleaved with either resistive plate chambers or streamer tubes~\cite{ifr}.

Signal and background ISR processes are simulated with 
Monte Carlo (MC) event generators based on
Ref.~\cite{EVA}. The  differential cross section for
$e^+e^-\to p\bar{p} \gamma$ is taken from Ref.~\cite{kuhn_pp}.
To analyze the experimental proton angular distribution, 
two samples of signal 
events are generated, one with $G_E=0$ and the other with $G_M=0$.
Since the polar-angle distribution of the ISR photon is peaked
near $0^\circ$ and $180^\circ$, the MC events are generated with
a restriction on the photon polar angle: 
$20^\circ<\theta_{\gamma}^\ast<160^\circ$ (the corresponding angular
range in the laboratory frame is $12^\circ <\theta_\gamma < 146^\circ$).
Extra soft-photon radiation from the initial state is generated
by the structure function method~\cite{strfun}. To restrict the
maximum energy of the extra photons, the
invariant mass of the hadron system and the ISR photon
is required to be greater than 8~GeV/$c^2$. 
For background
$e^+e^- \to \mu^+\mu^-\gamma$, $\pi^+\pi^- \gamma$, and $K^+K^-\gamma$
processes, final-state bremsstrahlung is generated using the  PHOTOS
package~\cite{PHOTOS}.
Background from $e^+e^- \to q\bar{q}$ is simulated with
the JETSET~\cite{JETSET} event generator; JETSET also
generates ISR events with hadron invariant mass above 2 GeV/$c^2$,
and therefore can be used to study ISR background with baryons
in the final state.
The dominant background process, $e^+e^- \to p\bar{p}\pi^0$,
is simulated separately by generating the angular and energy distributions 
for the final-state hadrons according to three-body phase space.

The response of the \babar\ detector is simulated using the 
Geant4~\cite{GEANT4} package.
The simulation takes into account the variations in the detector and
beam background conditions over the running period of the experiment.

\section{ \boldmath Event selection}
\label{selection}
The preliminary selection of $e^+e^- \to p\bar{p}\gamma$ candidates
requires that all of the final-state particles be detected inside a 
fiducial volume. Since a significant fraction of the events
contains beam-generated spurious tracks and photon candidates, we
select events with at least two tracks with opposite charge and
at least one photon candidate with $E^\ast_\gamma>3$~GeV.
The polar angle of the photon is required to be in the well-understood
region of the calorimeter:
$20^\circ <\theta_\gamma < 137.5^\circ$.
Each charged-particle track must originate from the interaction region, have
transverse momentum greater than 0.1~GeV/$c$ and be in the angular region
$25.8^\circ <\theta < 137.5^\circ$. The latter requirement
is needed to provide particle identification (PID) from the DIRC.
To suppress background from radiative Bhabha events, 
we reject events for which the ratios of calorimetric energy deposition to 
momentum for the two highest-momentum tracks satisfy the condition
$(E_{{\rm cal},1}/p_{1}-1)^2+(E_{{\rm cal},2}/p_{2}-1)^2<0.35^2$.

For events passing the preliminary selection, a kinematic fit is
performed to the $e^+e^- \to h^+h^- \gamma$ hypothesis with 
requirements of total energy and momentum conservation. 
Here $h$ can be $e$, $\mu$, $\pi$, $K$ or $p$, 
and $\gamma$ is the photon candidate with the
highest energy in the $e^+e^-$ c.m.~frame. 
For events with more than two charged tracks,
the fit uses the two oppositely charged tracks 
that pass closest to the interaction point. 
The MC simulation does not accurately reproduce the shape
of the photon energy resolution function. This leads to a
difference in the distributions of the $\chi^2$ of the kinematic
fit for data and simulated events. To reduce this difference, only
the measured direction of the ISR photon is used in the fit;
its energy is a free fit parameter.
For each of the five mass hypotheses,
the corrected angles and  energies of the particles and 
the $\chi^2$ value are obtained from the fit. 

The expected number of events from the background processes
$e^+e^-\to \pi^+ \pi^- \gamma$, $\mu^+\mu^- \gamma$, and $K^+K^- \gamma$ 
exceeds the number of signal events by two to three orders of magnitude. 
To suppress these backgrounds, we require that both charged particles be 
identified as protons according to the specific ionization $(dE/dx)$ 
measured in the SVT and DCH, and the Cherenkov angle measured
in the DIRC.  This requirement suppresses pion and muon
backgrounds by a factor of $3\times 10^4$, and kaon 
background by a factor $10^4$, with a loss of approximately 
30\% of the signal events.

\begin{figure}
\includegraphics[width=.4\textwidth]{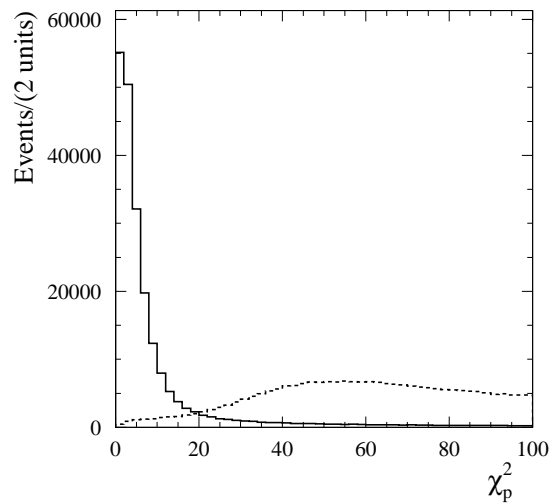}
\caption{The $\chi^2_p$ distribution for simulated
$e^+e^-\to p \bar{p}\gamma$ (solid histogram) and
$e^+e^-\to K^+K^-\gamma$ (dashed histogram, 
arbitrary normalization) events.
\label{fig2}}
\end{figure}

Background is further suppressed through requirements on the
$\chi^2$ of the kinematic fit:
$\chi^2_p<30$ and $\chi^2_K>30$, where  $\chi^2_p$ and
$\chi^2_K$ are the $\chi^2$ values of the kinematic fit for the
proton and kaon mass hypotheses, respectively.
The $\chi^2_p$ distribution for simulated 
$p\bar{p}\gamma$ events is shown in Fig.~\ref{fig2}. 
The long tail at high $\chi^2$ is due to 
events with extra soft photons emitted in the initial state. 
The dashed histogram is the $\chi^2_p$ distribution for 
$K^+K^-\gamma$ simulated  events. 
The $\chi^2$ requirements lead to the loss of 25\% of the signal events,
but provide additional background suppression by a factor of
50 for pion and muon events, and a factor of 30 for kaon events.

\begin{figure}
\includegraphics[width=.4\textwidth]{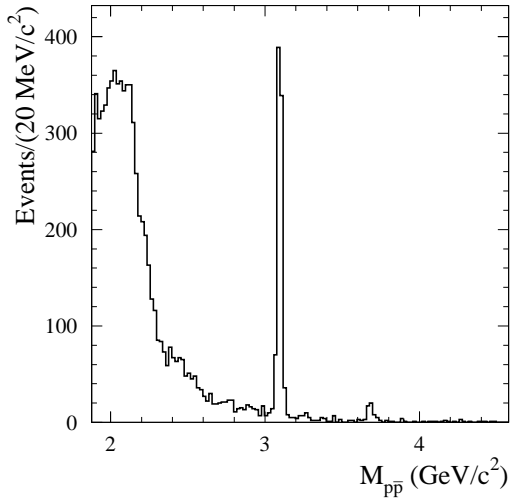}
\caption{The $p\bar{p}$ invariant mass spectrum for
the selected data $p\bar{p}\gamma$ candidates. The left edge
of the plot corresponds to the $p\bar{p}$ threshold.
\label{fig3}}
\end{figure}

The  $p\bar{p}$ invariant mass distribution 
is shown in Fig.~\ref{fig3}
for the 8298 data events that satisfy the selection criteria.
Most of the events have invariant mass below 3~GeV/$c^2$.
Signals from $J/\psi\to p\bar{p}$ and $\psi(2S)\to p\bar{p}$ 
decays are clearly seen.

\section{ \boldmath Background subtraction}
\label{subtraction}
Potential sources of background in the sample of 
selected $e^+e^-\to p\bar{p}\gamma$ candidates are the processes
$e^+e^-\to \pi^+\pi^-\gamma$,  $e^+e^-\to K^+K^-\gamma$,
$e^+e^-\to \mu^+\mu^-\gamma$, and $e^+e^-\to e^+e^-\gamma$
in which the charged particles are misidentified as protons.
Background contributions from processes with protons and neutral particle(s) 
in the final state, such as 
$e^+e^-\to p\bar{p}\pi^0$, $p\bar{p}\eta$, $p\bar{p}\pi^0\gamma$, {\it etc.},
are also anticipated.
 
Of particular interest is the possible background from the process 
$e^+e^-\to p\bar{p}\gamma$ with the photon emitted from the final state.
Due to the different charge-conjugation parity of the amplitudes corresponding to
initial-state radiation and final-state  radiation (FSR), 
their interference does not
contribute to the total $e^+e^-\to p\bar{p}\gamma$ cross section.
The contribution of the FSR amplitude is estimated to be~\cite{Chernyak}
$d\sigma/dm\approx |F_{\rm ax}|^2{8m\alpha^{3}\beta}/({27s^2})$,
where $F_{\rm ax}$ is the axial proton form factor.
Assuming $|F_{\rm ax}|\approx |G_M|$,  the ratio of the FSR and ISR cross 
sections is estimated to be about $10^{-3}$ for 
$p \bar{p}$ invariant masses below 4.5~GeV/$c^2$. We conclude that the FSR 
background is small and so may be neglected.
 
\subsection{\boldmath Background contributions from 
$e^+e^-\to \pi^+\pi^-\gamma$, $e^+e^-\to K^+ K^-\gamma$, 
$e^+e^-\to \mu^+\mu^-\gamma$ and $e^+e^-\to e^+e^-\gamma$}

 To estimate the background contribution from $e^+e^-\to \pi^+\pi^-\gamma$,
data and simulated $\pi^+\pi^-\gamma$ events are selected with the
following requirements on PID and on the $\chi^2$ of the kinematic fit:
\begin{enumerate}
\item one proton candidate, $\chi^2_{\pi}<20$;
\item one proton candidate, $\chi^2_{p}<30$, $\chi^2_{K}>30$;
\item two proton candidates, $\chi^2_{\pi}<20$;
\item two proton candidates, $\chi^2_{p}<30$, $\chi^2_{K}>30$.
\end{enumerate}
Here $\chi^2_{\pi}$ is the $\chi^2$ of the kinematic fit for
the pion mass hypothesis.
The fourth set of conditions corresponds to the standard selection criteria for
$p\bar{p}\gamma$ candidates.

\begin{figure}
\includegraphics[width=.4\textwidth]{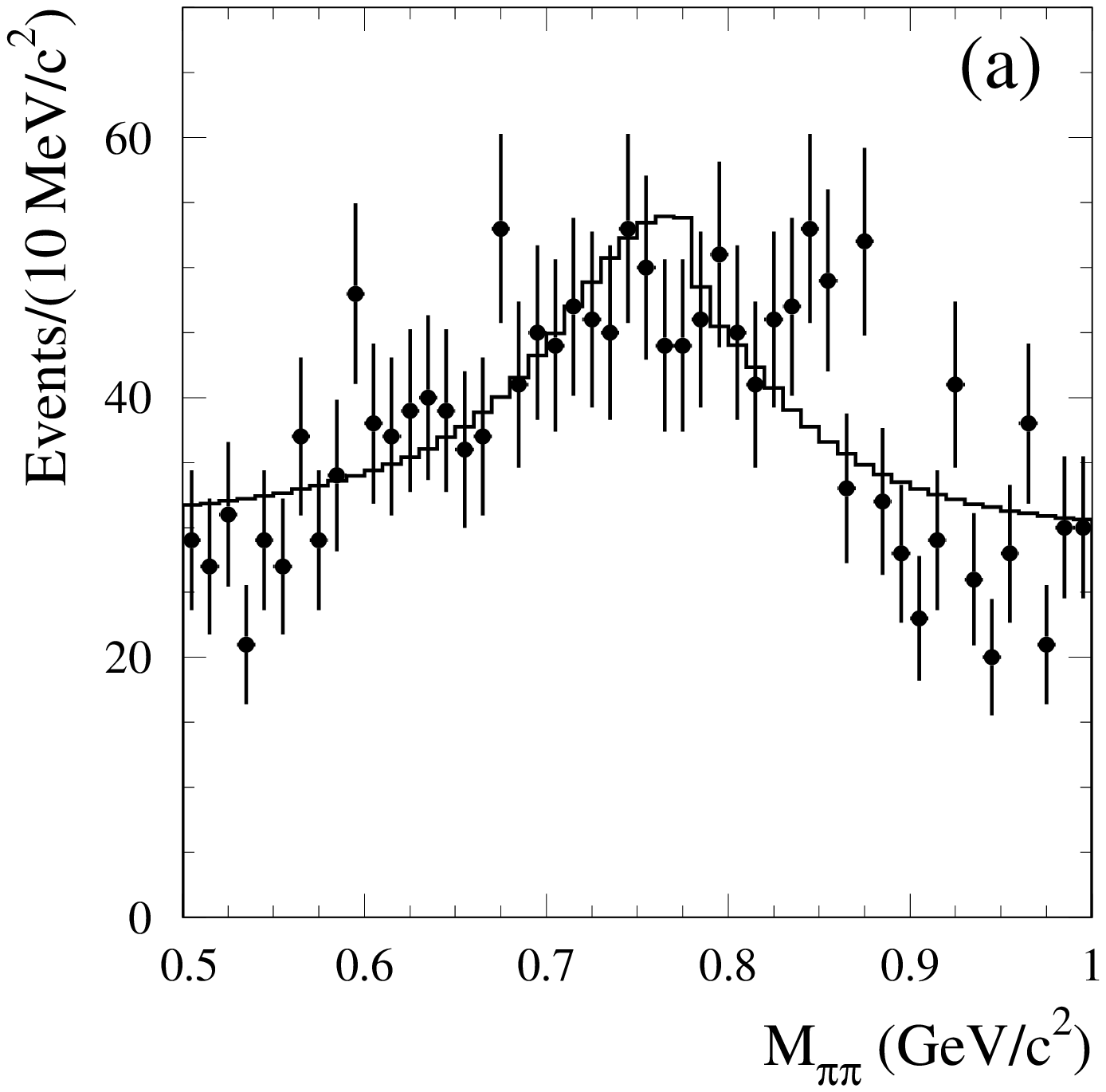}
\includegraphics[width=.4\textwidth]{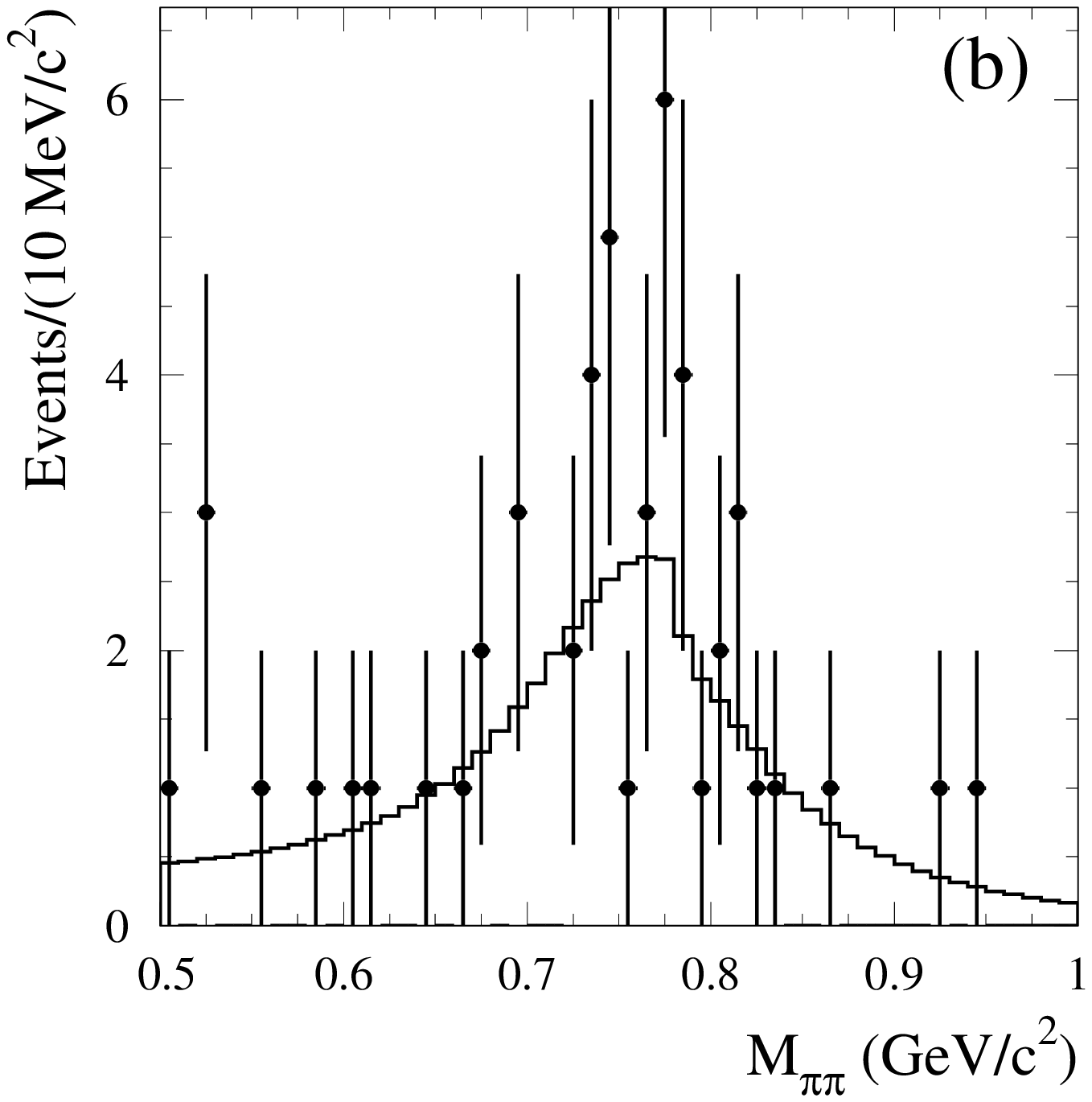}
\caption{(a) The $M_{\pi\pi}$ spectrum for data  events with $\chi^2_{p}<30$ 
and $\chi^2_{K}>30$, and one proton candidate (selection 2 in the text);
(b) the same spectrum for data events with $\chi^2_{\pi}<20$ and 
two proton candidates (selection 3 in the text). The histograms are the 
results of the fit described in the text.\label{fig4}}
\end{figure}

The invariant mass $M_{\pi\pi}$ of the two charged particles
under the pion-mass hypothesis is calculated;
the $M_{\pi\pi}$ distributions for data  selected
with criteria 2 and 3 are shown in Fig.~\ref{fig4}.
The data spectra are fit with a sum of the mass spectra
for simulated $\pi^+\pi^-\gamma$ events ($\rho$-meson line shape with 
$\omega$-$\rho$ interference) and 
a linear background term.  The numbers of $\pi\pi\gamma$ events 
with $0.5<M_{\pi\pi}<1$ GeV/$c^2$ obtained from the fits for selections
1--3 are listed in Table~\ref{pibkg_tab}, together with the corresponding 
numbers from the $\pi^+\pi^-\gamma$ MC simulation. The spectrum for
selection 4 is fit by a linear function; no $\rho$-meson
contribution is needed to describe this spectrum. 
\begin{table}
\caption{The numbers of $\pi\pi\gamma$ events for data and MC simulation
with $0.5<M_{\pi\pi}<1$~GeV/$c^2$ that satisfy  different selection
criteria for data and MC simulation.
The data numbers are obtained from the fits to the $M_{\pi\pi}$ distributions
described in the text.
\label{pibkg_tab}}
\begin{ruledtabular}
\begin{tabular}{ccc}
selection & data & MC \\
\hline
\\[-2.1ex]
1&$15310\pm160$&$14800\pm180$\\
2&$400\pm60$&$460\pm30$\\
3&$41\pm8$&$48\pm11$\\
\end{tabular}
\end{ruledtabular}
\end{table}

Since the simulation correctly predicts the numbers of pion events
for selections 1--3, it can be used to estimate the pion
background for the standard selection 4.
We observe no events satisfying the standard selection criteria
in the $\pi\pi\gamma$ MC sample. The corresponding upper limit on 
$\pi\pi\gamma$ background in the data sample is 5.2 events at 90\% confidence
level (CL).
The estimated pion background is less than 0.1\% of the number of selected
$p\bar{p}\gamma$ candidates.

Similarly, the number of $e^+e^-\to K^+K^-\gamma$ events can be estimated
from the number of events in the $\phi$ meson peak in the 
distribution of invariant mass of the charged particles
calculated under the kaon hypothesis.
It is found that the $K^+K^-\gamma$ MC simulation predicts reasonably
well the numbers of kaon events in the data sample with
one identified kaon and the standard $\chi^2$ conditions,
and in the data sample with two identified kaons and 
$\chi^2_K<20$. Therefore we use
the MC simulation to estimate kaon background for
the standard selection. The estimated background, $1.6\pm0.8$
events, is significantly less than 0.1\% of the number of
data events selected.

The kinematic properties of the $e^+e^-\to e^+ e^-\gamma$ process are used
to estimate the electron background. About 50\% of $e^+ e^-\gamma$ events have
$e^+e^-$ invariant mass between 3 and 7~GeV/$c^2$ and
$\cos{\psi^\ast}< -0.98$, where  $\psi^\ast$ is the
angle between the two tracks in the initial $e^+e^-$  c.m.~frame.
In the event sample with two proton candidates we do not find
events having the above characteristics.
The corresponding  90\% CL upper limit on the $e^+ e^-\gamma$ background in
the data sample is 4.6 events (2 events with $M_{p\bar{p}}<4.5$ GeV/$c^2$).

To compare MC simulation and data for the process $e^+e^-\to\mu^+\mu^-\gamma$,
we use a subsample of events selected with the requirement that
both charged particles be identified as muons.
Muon identification is based on IFR information, and does not
use DIRC or $dE/dx$ information, which are necessary for proton
identification. 
In the data samples with one or two identified protons 
obtained with the standard $\chi^2$ selection, we select 86 and 2 
muon-identified events, respectively.
These numbers can be compared with $60\pm16$ and zero events expected from 
the $e^+e^-\to\mu^+\mu^-\gamma$ simulation. 
Taking into account that the ratio of the total number of $\mu^+\mu^-\gamma$
events to those with two identified muons is about two-to-one, we estimate
the $\mu^+\mu^-\gamma$ background for the standard selection criteria
to be $4.0\pm 2.8$ events.

The combined background from the processes 
$e^+e^-\to C^+C^-\gamma$, $C=\pi,K,e,\mu$
is less than 0.2\% of the number of selected
$p\bar{p}\gamma$ candidates, and so can be neglected.

\subsection{Background from $e^+e^-\to p\bar{p}\pi^0$\label{pppi}}
\begin{figure}
\includegraphics[width=.40\textwidth]{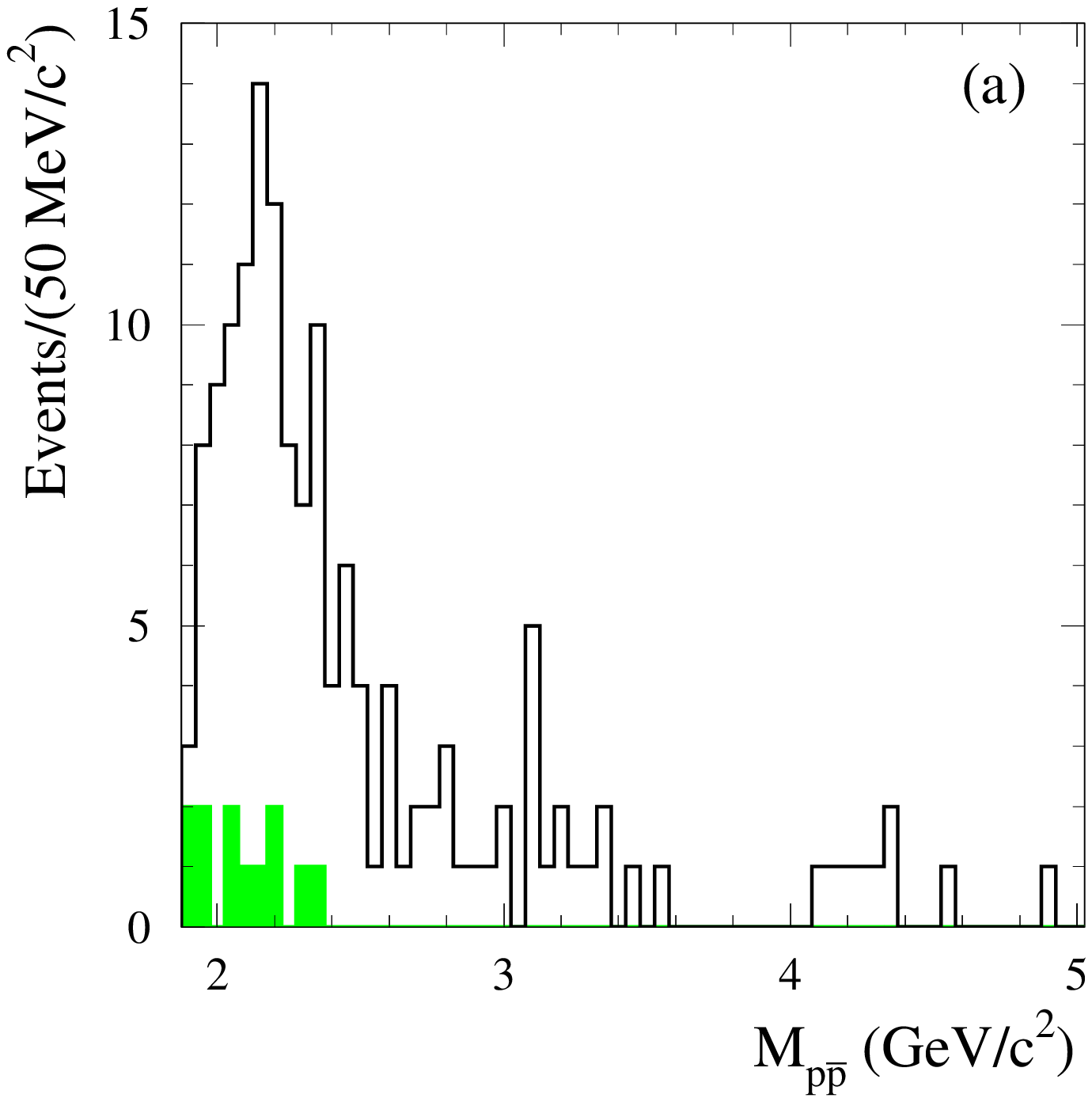}
\includegraphics[width=.40\textwidth]{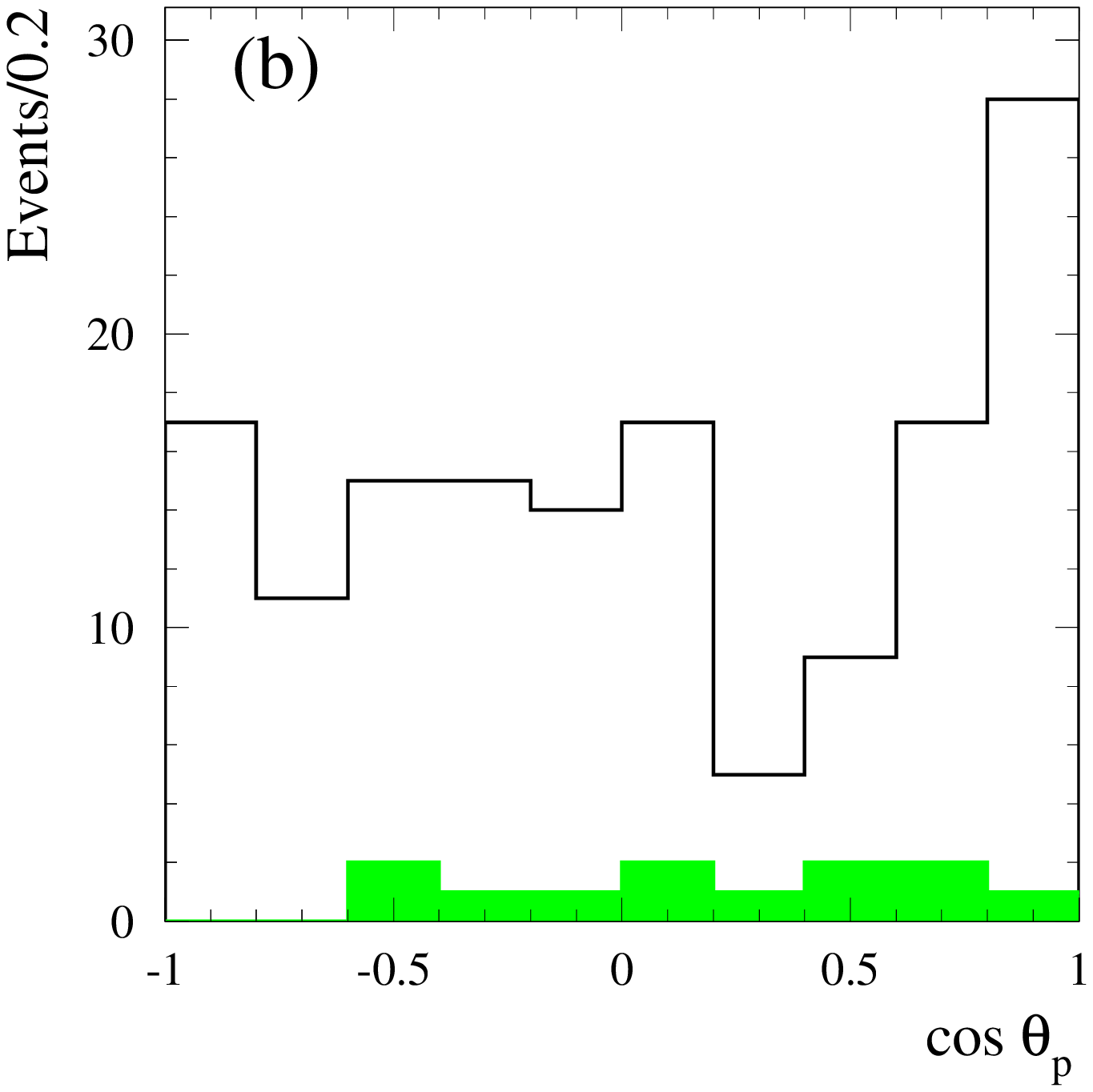}
\caption{(a) The $M_{p\bar{p}}$  spectrum and (b)
the $\cos{\theta_{p}}$ distribution for selected
$e^+e^-\to p\bar{p}\pi^0$ candidates in data. In each figure, the 
shaded histogram shows the background contribution estimated from 
the $M_{\gamma\gamma}$ sidebands.
\label{fig5}}
\end{figure}
The dominant source of background to the $e^+e^-\to p\bar{p} \gamma$ process
arises from  $e^+e^-\to p\bar{p}\pi^0$.  A significant fraction 
of $p\bar{p}\pi^0$ events with an undetected low-energy
photon, or with merged photons from the $\pi^0$ decay, is reconstructed
under the $p\bar{p}\gamma$ hypothesis with a low value of $\chi^2$, 
and so cannot be separated from the process under study. 
This background is studied by selecting a special subsample of
events containing two charged particles identified as protons
and at least two photons with energy greater than 0.1~GeV,  one of
which must have c.m.~energy above 3~GeV.
   The two-photon invariant mass is required to be in the
range 0.07--0.20~GeV/$c^2$, which is centered on the nominal $\pi^0$ mass. 
A kinematic fit to the
$e^+e^-\to p\bar{p}\gamma\gamma$ hypothesis is then performed.
Requirements on the $\chi^2$ of the kinematic fit ($\chi^2<25$)
and the two-photon invariant mass ($0.1025<M_{\gamma\gamma}<0.1675$~GeV/$c^2$)
are imposed in order to select $e^+e^-\to p\bar{p}\pi^0$ candidates.
The $M_{\gamma\gamma}$ sidebands
$0.0700<M_{\gamma\gamma}<0.1025$~GeV/$c^2$ and
$0.1675<M_{\gamma\gamma}<0.2000$~GeV/$c^2$ are used to
estimate background.
The $M_{p\bar{p}}$  spectra and $\cos{\theta_{p}}$ distributions
for data events from the signal and sideband $M_{\gamma\gamma}$ regions 
are shown in Fig.~\ref{fig5}.
The total number of selected events is 148 in the signal region and
12 in the sidebands. The number of $e^+e^-\to p\bar{p}\pi^0$ events 
in the $M_{\gamma\gamma}$ sidebands expected from MC simulation is 5.4.

\begin{figure}
\includegraphics[width=.4\textwidth]{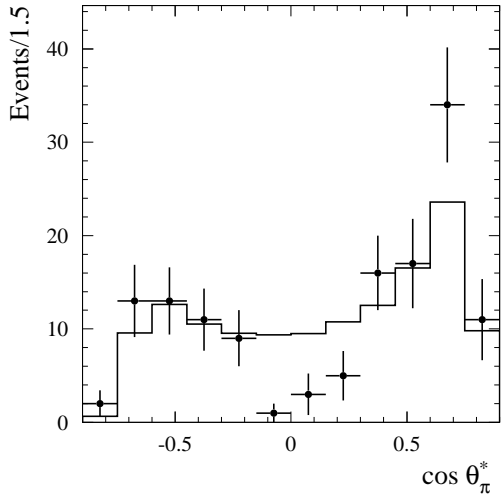}
\caption{The $\cos{\theta_{\pi}^\ast}$ distribution for 
$e^+e^-\to p\bar{p}\pi^0$ event candidates for data (points with error bars) 
and simulation (histogram).
\label{fig6}}
\end{figure}
The $p\bar{p}\pi^0$ selection criteria described above
are applied to simulated $e^+e^-\to q\bar{q}$ events generated with
the JETSET program.
In the mass region $M_{p\bar{p}}<5$ GeV/$c^2$, the predicted number 
of $e^+e^-\to p\bar{p}\pi^0$ events is $120\pm9$.
These events exhibit an
enhancement in the $M_{p\bar{p}}$ distribution near the $p\bar{p}$ threshold,
similar to that in data.
However, the $\cos{\theta_{p}}$ distribution of MC events is peaked near
$\cos{\theta_{p}}=\pm1$, whereas for the data the distribution is flat.
To study the $e^+e^-\to p\bar{p}\pi^0$ background, the sample of
simulated $e^+e^-\to p\bar{p}\pi^0$
events is generated according to three-body phase space, but with
an additional weight proportional to $(M_{p\bar{p}}-2m_p)^{{3}/{2}}$
to imitate the $M_{p\bar{p}}$ distribution observed in
data. The resulting generated $\cos{\theta_{p}}$  distribution is flat.

\begin{figure}
\includegraphics[width=.4\textwidth]{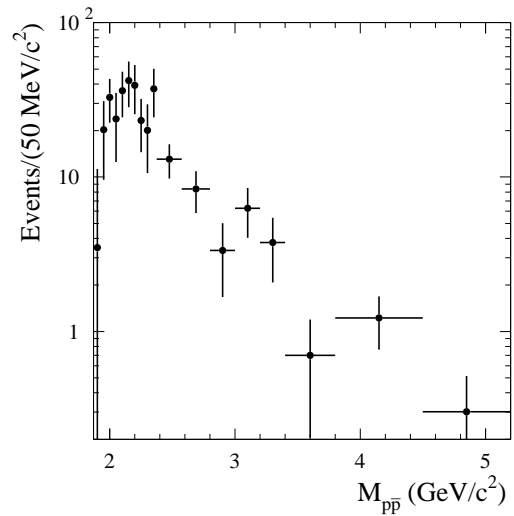}
\caption{The expected $M_{p\bar{p}}$ spectrum for
$e^+e^-\to p\bar{p}\pi^0$ events selected with the 
standard $p\bar{p}\gamma$
criteria. The spectrum is obtained by scaling the data distribution
shown in Fig.~\ref{fig5}(a)    
by the factor $K_{\rm MC}(M_{p\bar{p}})$ described in the text.
\label{fig7}}
\end{figure}
In Fig.~\ref{fig6} the $\cos{\theta_{\pi}^\ast}$ distribution
for selected data and simulated $e^+e^-\to p\bar{p}\pi^0$ events is shown,
where $\theta_{\pi}^\ast$ is the $\pi^0$ polar angle in the $e^+e^-$ c.m.~frame.
It is seen that the data and simulated distributions differ slightly.
Since we do not observe a significant variation of the $\cos\theta_{\pi}^\ast$
distribution with $M_{p\bar{p}}$ in data, we use the data distribution
averaged over $M_{p\bar{p}}$ (Fig.~\ref{fig6}) to reweight the
$e^+e^-\to p\bar{p}\pi^0$ simulation.
\begin{table*}
\caption{ The number of selected $p\bar{p}\gamma$
candidates, $N_{p\bar{p}\gamma}$, and the number of background events from
the $e^+e^-\to p\bar{p}\pi^0$ process, $N_{p\bar{p}\pi^0}$, for different
ranges of $M_{p\bar{p}}$. The $p\bar{p}$ mass ranges near the $J/\psi$ and 
$\psi(2S)$ resonances are excluded.
\label{pppi0_tab}}
\begin{ruledtabular}
\begin{tabular}{lccccc}
$M_{p\bar{p}}$ (GeV/$c^2$) & $<2.50$  & 2.50--3.05 & 3.15--3.60 & 3.75--4.50 & $>4.5$ \\[0.3ex]
\hline
\\[-2.1ex]
$N_{p\bar{p}\gamma}$& 6695     &   592    &    76    &   29    &    9   \\
$N_{p\bar{p}\pi^0}$ &$321\pm37$&$66\pm15$ & $26\pm9$ & $17\pm6$ & $6\pm3$ \\
\end{tabular}
\end{ruledtabular}
\end{table*}
\begin{table*}
\caption{The number of selected $p\bar{p}\gamma$ 
candidates from the mass region $M_{p\bar{p}}< 4.5$ GeV/$c^2$ with 
$\chi^2_{p}<30$ ($N_1$) and $30<\chi^2_{p}<60$ ($N_2$) for signal and
for different background processes; $\beta_i$ is the ratio $N_2/N_1$ 
obtained from simulation. The first column shows the numbers of 
$p\bar{p}\gamma$ candidates selected in data.
The numbers for $e^+e^-\to p\bar{p}\gamma$ are obtained from data 
using the background subtraction procedure described in the text.
 \label{beta}}
\begin{ruledtabular}
\begin{tabular}{lccccccc}
&data&$p\bar{p}\pi^0$&$e^+e^-$&Other ISR&$p\bar{p}\gamma$\\[0.3ex]
\hline
\\[-2.1ex]
$N_1$ &8298&$448\pm42$&$40\pm5$&$55\pm6$&$7741\pm113$\\
$N_2$ &560 &$79\pm7$  &$76\pm7$&$74\pm7$&  $337\pm16$\\
$\beta_i$&&$0.175\pm0.04$&$1.88\pm0.29$&$1.34\pm0.18$&$0.0435\pm0.0020$\\
\end{tabular}
\end{ruledtabular}
\end{table*}

From the reweighted simulation, we calculate the ratio ($K_{\rm MC}$) 
of the $M_{p\bar{p}}$ distribution for events selected with the
standard $p\bar{p}\gamma$ criteria to that selected with 
the $p\bar{p}\pi^0$ criteria.
The value of the ratio $K_{\rm MC}(M_{p\bar{p}})$ varies from
3.4 near the $M_{p\bar{p}}$ threshold to 2.0 at 5~GeV/$c^2$.
The expected $M_{p\bar{p}}$ spectrum for the $e^+e^-\to p\bar{p}\pi^0$
 background satisfying the $p\bar{p}\gamma$ selection criteria is
shown in Fig.~\ref{fig7}, and is evaluated as
$K_{\rm MC}(M_{p\bar{p}}) \times(dN/dM_{p\bar{p}})_{data}$,
where $(dN/dM_{p\bar{p}})_{data}$ is the mass distribution for
$e^+e^-\to p\bar{p}\pi^0$ events obtained above (Fig.~\ref{fig5}(a)).
The number of selected $e^+e^-\to p\bar{p}\gamma$
candidates and the expected number of $e^+e^-\to p\bar{p}\pi^0$ background 
events are given for different $p\bar{p}$ mass ranges in Table~\ref{pppi0_tab}.
The background contribution  grows from 5\% near $p\bar{p}$ threshold to 50\%
at $M_{p\bar{p}}\approx 4$~GeV/$c^2$. All observed $p\bar{p}\gamma$ candidates
with $M_{p\bar{p}}> 4.5$~GeV/$c^2$ are consistent with $p\bar{p}\pi^0$ 
background.

\subsection{Other sources of background}
Other possible background sources are ISR processes with higher final-state
multiplicity ($e^+e^-\to p\bar{p}\pi^0\gamma$, $p\bar{p}\,2\pi^0\gamma$, 
{\ldots}), and direct $e^+e^-$ annihilation processes other than
$e^+e^-\to p\bar{p}\pi^0$ ($e^+e^-\to p\bar{p}\eta$,
$e^+e^-\to p\bar{p}\,2\pi^0$, and so on). All of these processes are 
simulated by JETSET. The simulation leads to the prediction that the ISR 
background is $55\pm6$ events, and that the direct annihilation background is 
$40\pm5$ events. The total predicted background from these two sources is 
about 1.2\% of the number of selected $p\bar{p}\gamma$ candidates. We do not 
perform a detailed study of these background processes. Their contribution 
is estimated from data by using the $\chi^2$ sideband region, 
as described below in Sec.~\ref{bkgsub}.

\subsection{Background subtraction}\label{bkgsub}
Table~\ref{beta} summarizes the expected number of background events estimated
in the previous sections. The ``other ISR'' and ``$e^+e^-$'' columns show the 
background contributions estimated with JETSET that result from ISR processes,
and from $e^+e^-$ annihilation processes other than $e^+e^-\to p\bar{p}\pi^0$.
Because JETSET has not been verified precisely  
for the rare processes contributing to the $p\bar{p}\gamma$
candidate sample, we use a method of background estimation that is 
based on the difference in $\chi^2$ distributions
for signal and background events. 
The first and second rows in Table~\ref{beta}
show the expected numbers of signal and background events
with $\chi^2_{p}<30$ ($N_1$) and $30<\chi^2_{p}<60$ ($N_2$).
The last row lists $\beta_i$, the ratio of $N_2$ to $N_1$.
From Table~\ref{beta}, it is evident that 
the coefficient $\beta_i$ 
for signal events, and for background events from the
processes with higher hadron multiplicity 
(``$e^+e^-$'' and ``Other ISR'' columns), are very different. 
This difference can be used
to estimate the background from these two sources, as follows.
First,  the $p\bar{p}\pi^0$ background 
determined as described in Sec.~\ref{pppi}
is subtracted from data.
Then, from the corrected numbers of events in the signal and sideband
$\chi^2$ regions, $N_1^\prime$ and $N_2^\prime$, the 
numbers of signal and background (from ``$e^+e^-$'' and ``ISR'' sources)
events with $\chi^2_{p}<30$ can be calculated:
\begin{eqnarray}
\label{bkgsub_eq}
N_{sig}&=&\frac{N_1^\prime-N_2^\prime/\beta_{bkg}}
{1-\beta_{p\bar{p}\gamma}/\beta_{bkg}},\\
N_{bkg}&=&N_1^\prime-N_{sig},\nonumber
\end{eqnarray}
where $\beta_{bkg}$ is the ratio of the fractions of events 
in the sideband and signal $\chi^2$ regions
averaged over all background processes of the ``$e^+e^-$'' and ``ISR'' types.
For this coefficient, $\beta_{bkg}=1.6\pm0.3$ is used;
it is the  average of
$\beta_{e^+e^-}$ and $\beta_{\rm ISR}$ with the uncertainty 
$(\beta_{e^+e^-}-\beta_{\rm ISR})/2$.

        In Table~\ref{beta},  it is also evident
that $p\bar{p}\gamma$ events dominate the $\chi^2$ sideband region.
Therefore, the  background is very sensitive
to the accuracy of the $\beta_{p\bar{p}\gamma}$ coefficient. In particular, the
data-Monte Carlo difference in the $\chi^2$ distribution can lead to 
a systematic shift of the result. The simulation of the $\chi^2$ distribution
for $p\bar{p}\gamma$ events is validated using data and simulated 
$e^+e^-\to \mu^+\mu^-\gamma$ events. These are
very similar kinematically to the process under study, and can be 
selected with negligible background.
The ratio of the $\beta $ coefficients for $e^+e^-\to \mu^+\mu^-\gamma$ 
data and simulation is independent of the $\mu^+\mu^-$ mass and is equal to
$1.008\pm0.008$. This ratio is used to correct the $\beta_{p\bar{p}\gamma}$
value obtained from simulation, which varies from 0.043 at $p\bar{p}$
threshold to 0.048 at 4.5 GeV/$c^2$. 

With the method described above,  the total numbers
of $e^+e^-\to p\bar{p}\gamma$ events ($N_{sig}$) and 
background events from ``$e^+e^-$'' and ``ISR'' sources ($N_{bkg}$)
in the signal region are found to be $7741 \pm 95 \pm 62$ and 
$109\pm16\pm25$, respectively.
The main source of the systematic uncertainty on $N_{sig}$ is
the uncertainty in the $p\bar{p}\pi^0$ background. The number
of background events is in good agreement with the estimate 
from simulation, $(40\pm5) + (55\pm6)=95\pm8$. The total background in the
$\chi^2_{p}<30$ region is $531\pm51$ events, which is about 7\%
of the number of signal events.  

The background subtraction procedure is performed in each $p\bar{p}$ mass 
interval. 
The number of selected events for each interval after background subtraction
and unfolding event migration between intervals (see Sec.~\ref{xsec})
is listed in Table~\ref{sumtab}.
The events from $J/\psi$ and $\psi(2S)$ decays are subtracted from the
contents of the corresponding intervals (see Sec.~\ref{jpsi}).

\section{Angular distributions}
The modulus of the ratio of the electric and magnetic
form factors can be extracted from an analysis of the distribution 
of $\theta_p$, the angle between the proton momentum in
the $p\bar{p}$ rest frame and the momentum of the $p\bar{p}$ system
in the $e^+e^-$ c.m.~frame. 
This distribution is given by
\begin{eqnarray}
\lefteqn{\frac{dN}{d\cos{\theta_p}}=} \nonumber \\
&A\left(H_M(\cos{\theta_p},M_{p\bar{p}})+
\left |\frac{G_E}{G_M}\right|^2H_E(\cos{\theta_p},M_{p\bar{p}})\right).
\label{an_fit}
\end{eqnarray}
The functions $H_M(\cos{\theta_p},M_{p\bar{p}})$ and 
$H_E(\cos{\theta_p},M_{p\bar{p}})$
do not have an analytic form, and so are determined using MC simulation.
To do this, two samples of $e^+e^-\to p\bar{p}\gamma$ events are generated,
one with $G_E=0$ and the other with $G_M=0$. The functions obtained are 
close to the  $1+\cos^2 \theta_p$ and $\sin^{2}\theta_p$ functions
describing angular distributions for the magnetic and electric form factors
in the case of $e^+e^-\to p\bar{p}$.

\begin{figure*}
\includegraphics[width=0.40\textwidth]{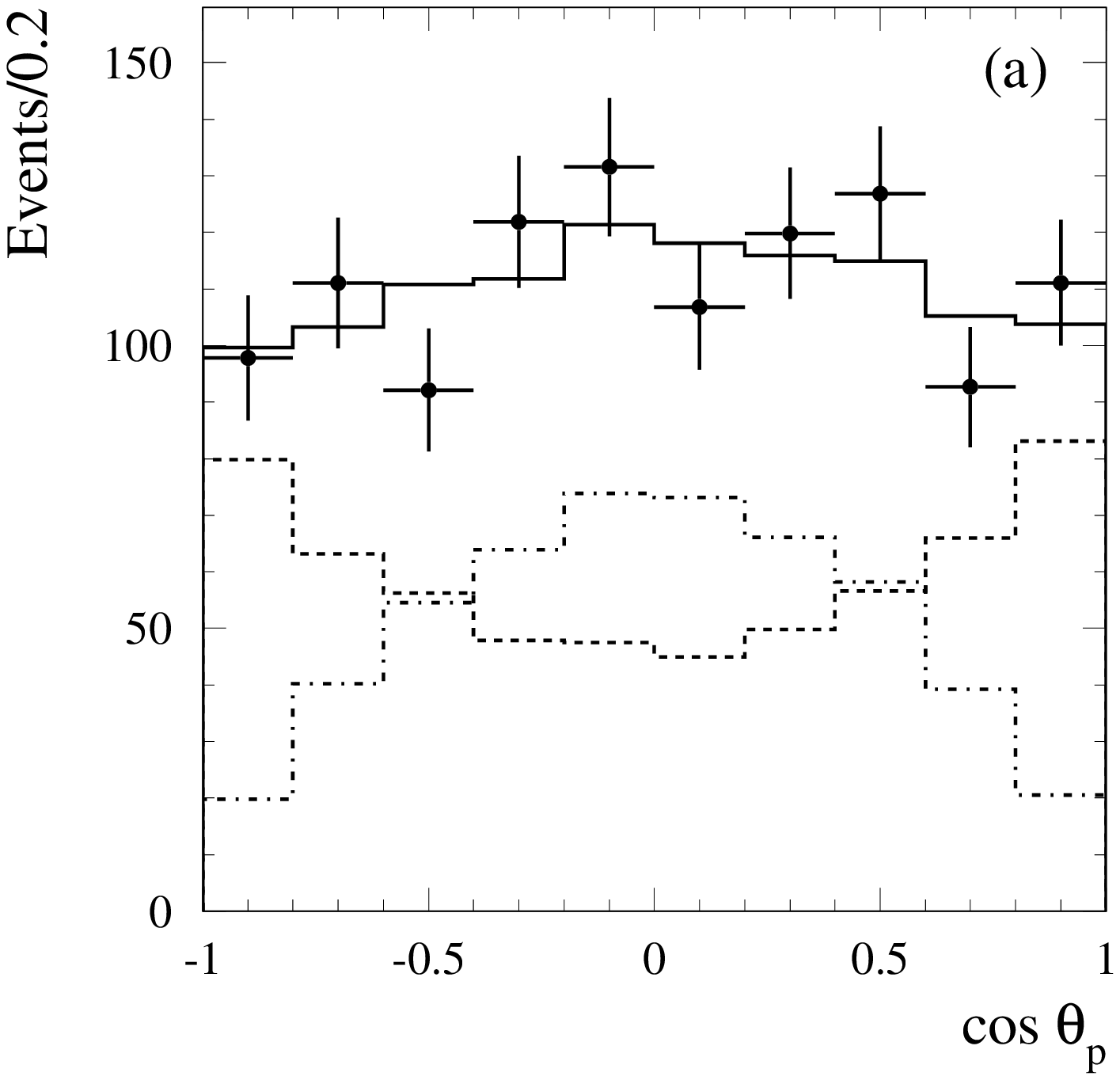}
\includegraphics[width=0.40\textwidth]{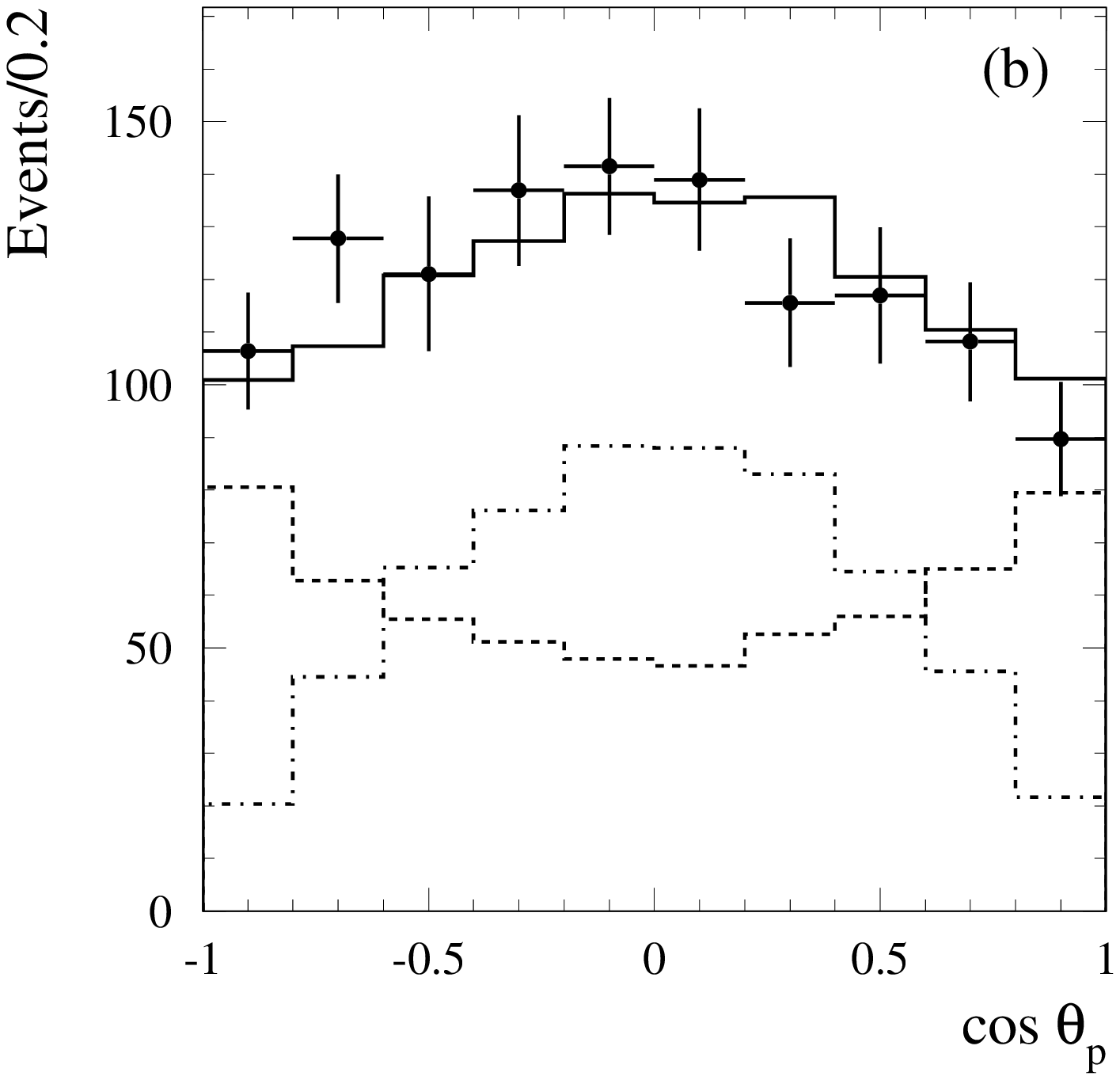}
\vspace{2mm}\\
\includegraphics[width=0.40\textwidth]{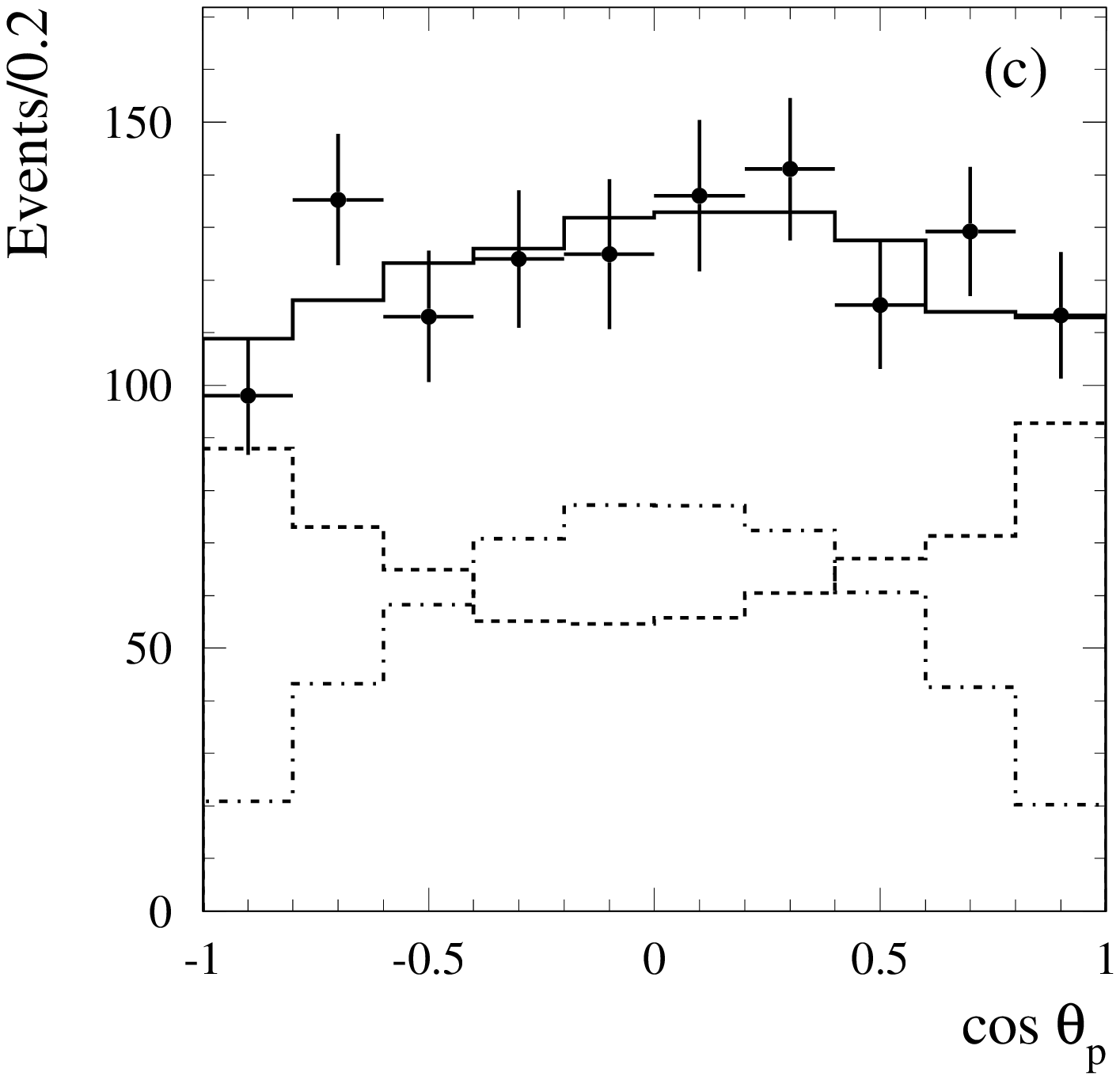}
\includegraphics[width=0.40\textwidth]{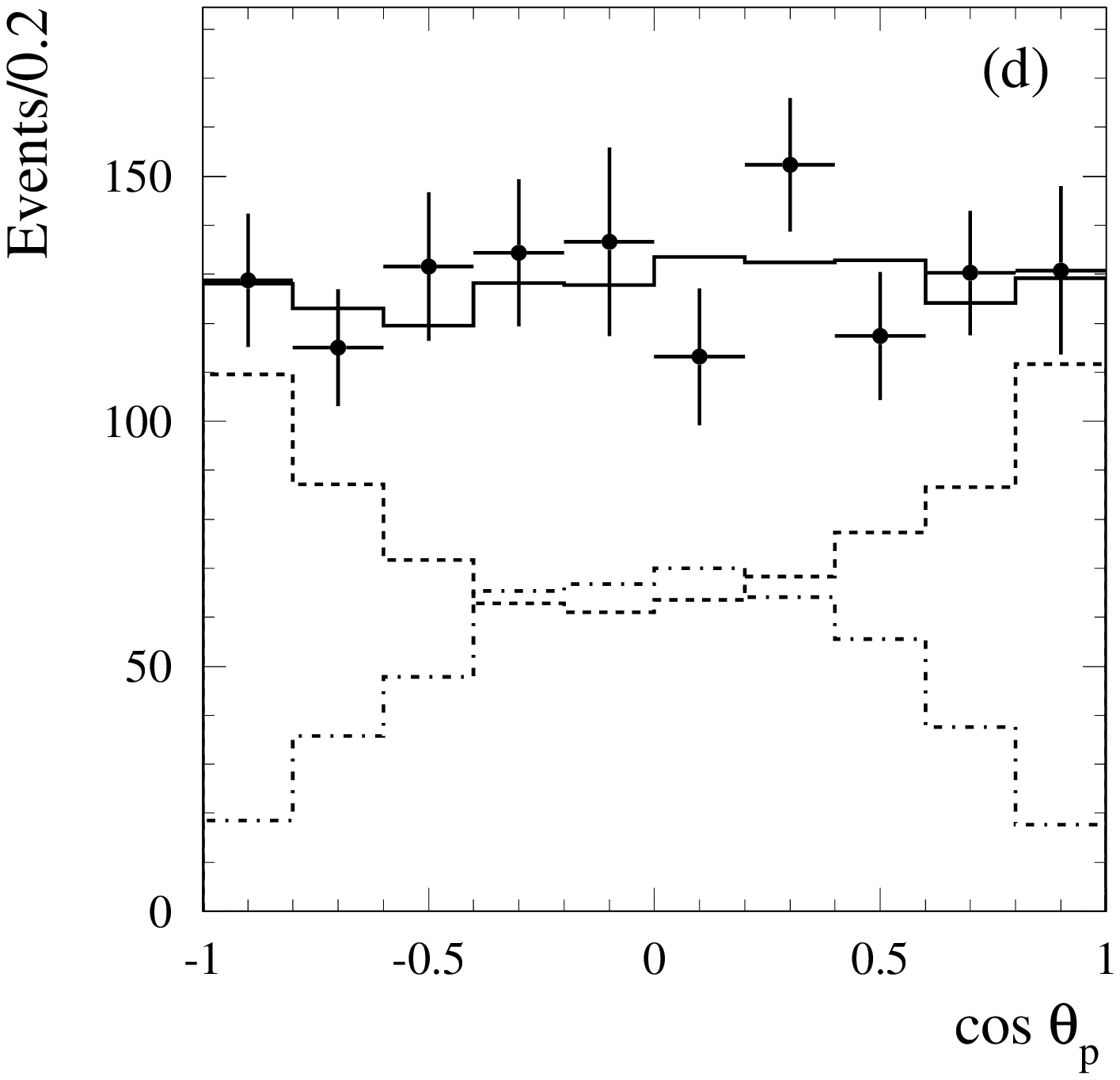}
\vspace{2mm}\\
\includegraphics[width=0.40\textwidth]{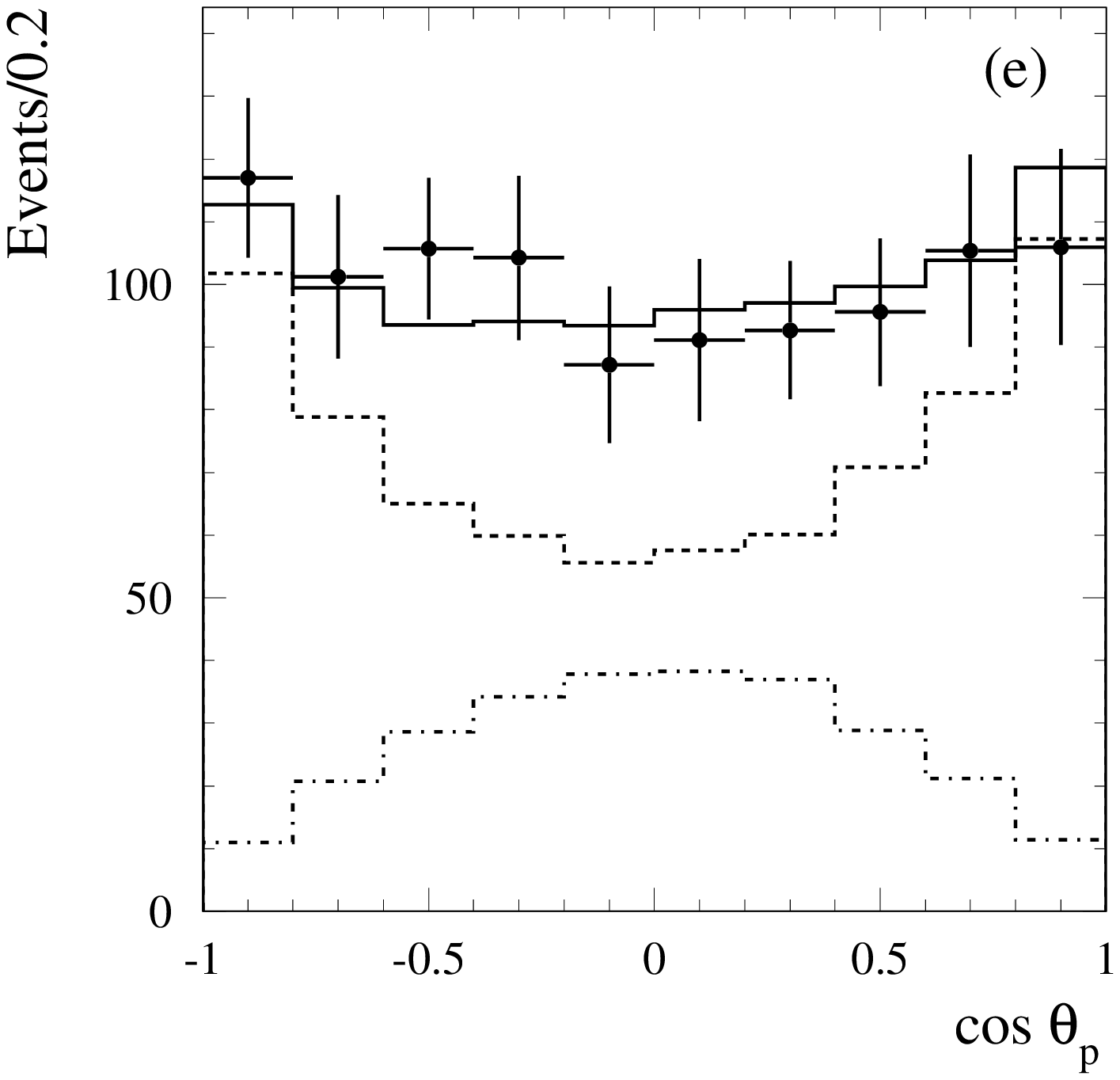}
\includegraphics[width=0.40\textwidth]{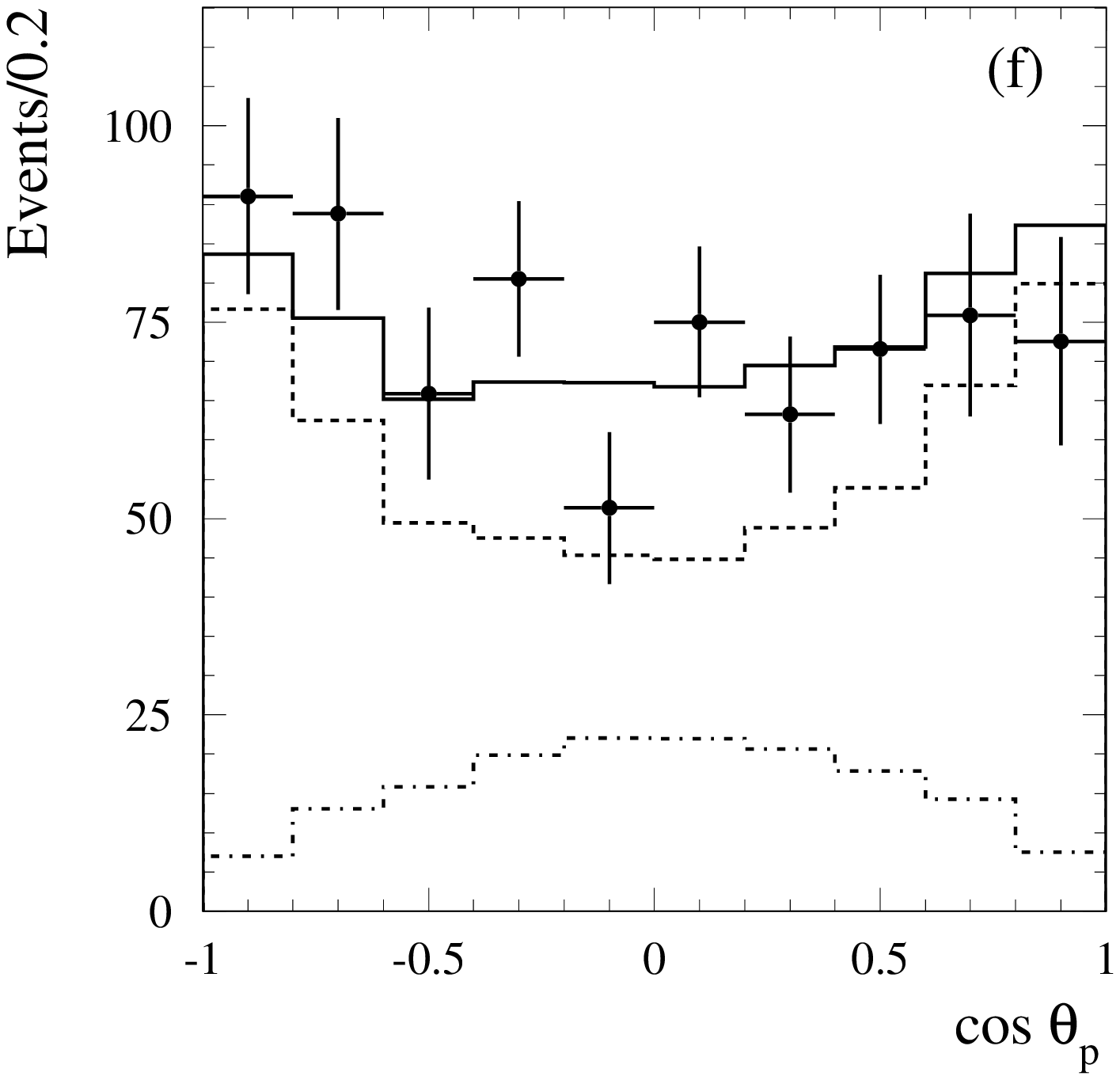}
\caption{The $\cos{\theta_p}$ distributions for
different $p \bar{p}$ mass regions:
(a) 1.877--1.950~GeV/$c^2$,
(b) 1.950--2.025~GeV/$c^2$,
(c) 2.025--2.100~GeV/$c^2$,
(d) 2.100--2.200~GeV/$c^2$,
(e) 2.200--2.400~GeV/$c^2$,
(f) 2.400--3.000~GeV/$c^2$.
The points with error bars show the data distributions after
background subtraction.  
The histograms result from the fits: the dashed histograms
correspond
to the magnetic form factor contributions and the dot-dashed histograms
to the electric form factor contributions.
\label{fig8}}
\end{figure*}

The observed angular distributions are fit in 
six ranges of $p \bar{p}$ invariant mass
from threshold to 3~GeV/$c^2$.
The fit intervals, the corresponding numbers of selected
events, and  the estimated numbers of background events are listed
in Table~\ref{ang_tab}.
\begin{table}
\caption{The number of selected $p\bar{p}\gamma$
candidates ($N$) and the number of background events ($N_{bkg}$) for
each $p\bar{p}$ mass interval; $|G_E/G_M|$ is the fitted ratio of form 
factors.\label{ang_tab}}
\begin{ruledtabular}
\begin{tabular}{cccc}
$M_{p\bar{p}}$, GeV/$c^2$ & $N$  & $N_{bkg}$ & $|G_E/G_M|$ \\[0.3ex]
\hline
\\[-2.1ex]
1.877--1.950        & 1162  & $19\pm10 $ & $1.36_{-0.14-0.04}^{+0.15+0.05}$ \\
\\[-2.1ex]
1.950--2.025        & 1290  & $53\pm16$ & $1.48_{-0.14-0.05}^{+0.16+0.06}$ \\
\\[-2.1ex]
2.025--2.100        & 1328  & $63\pm14$ & $1.39_{-0.14-0.07}^{+0.15+0.07}$ \\
\\[-2.1ex]
2.100--2.200        & 1444  & $118\pm28$ & $1.26_{-0.13-0.09}^{+0.14+0.10}$ \\
\\[-2.1ex]
2.200--2.400        & 1160  & $126\pm26$ & $1.04_{-0.16-0.10}^{+0.16+0.10}$ \\
\\[-2.1ex]
2.400--3.000        & 879   & $122\pm22$ & $1.04_{-0.25-0.15}^{+0.24+0.15}$ \\
\\[-2.1ex]
\end{tabular}
\end{ruledtabular}
\end{table}
For each $p\bar{p}$ mass interval and each angular interval the background 
is subtracted  using the procedure described in Section \ref{bkgsub}. 
The angular distributions obtained are shown in Fig.~\ref{fig8}.
The distributions are fit to Eq.~(\ref{an_fit}) with
two free parameters: $A$ (the overall normalization)
and $|G_E/G_M|$. The
functions $H_M$ and $H_E$ are replaced by the histograms
obtained from MC simulation with the $p\bar{p}\gamma$ selection
criteria applied.
To account for differences between the $p\bar{p}$ mass distributions
of $p\bar{p}\gamma$ events in data and  MC simulation,
the histograms $H_M$ and $H_E$ are recalculated using weighted
events.
The weights are obtained from the ratio
of the $p\bar{p}$ mass distributions in data and simulation.
In principle,  the weights for $H_M$ and $H_E$ differ
due to the different mass dependences of $G_M$ and $G_E$.  
A first approximation uses $G_M=G_E$. The fitted values of
$|G_E/G_M|$ are then used in the next approximation
to recalculate $H_M$ and $H_E$. The second iteration
leads to a small change (less than 2\%) in the fitted values,
and the procedure converges after a third iteration.

\begin{figure}[h]
\includegraphics[width=.4\textwidth]{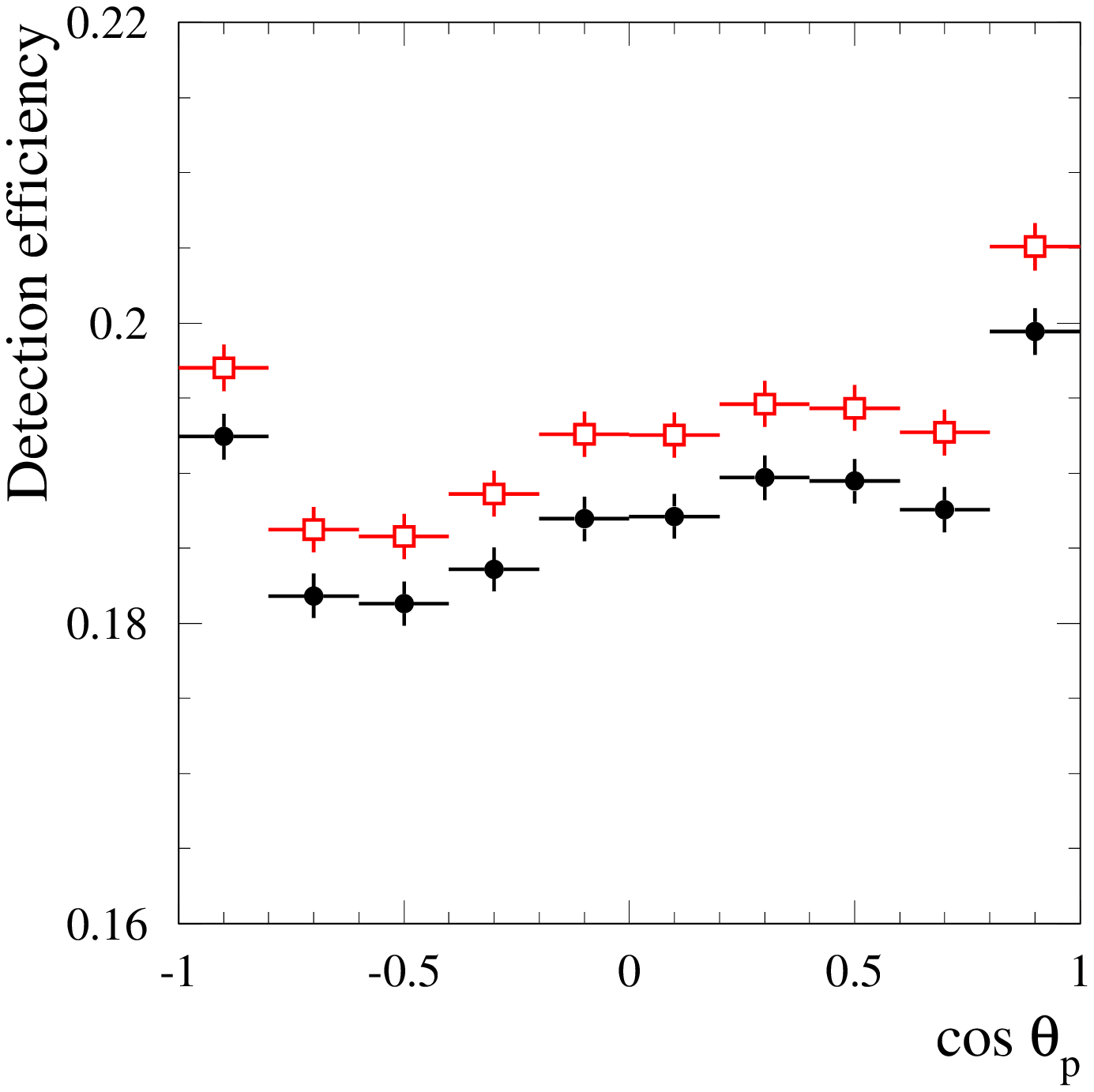}
\caption{The angular dependence of the detection efficiency
for simulated events with $M_{p\bar{p}}<2.5$~GeV/$c^2$
before (open squares) and after (filled circles) correction
for data-simulation differences in detector response.
\label{fig9}}
\end{figure}
The simulated angular distributions are corrected to
account for the differences between data and simulation in
particle identification, tracking, and photon efficiency. These corrections
are discussed in detail in the next section.
The angular dependence of the detection efficiency calculated
with MC simulation before and after the corrections
is shown in Fig.~\ref{fig9}.
The deviations from uniform efficiency, which do not exceed 10\%,
arise from the momentum dependence of proton/antiproton 
particle identification efficiency.

\begin{figure}
\includegraphics[width=.4\textwidth]{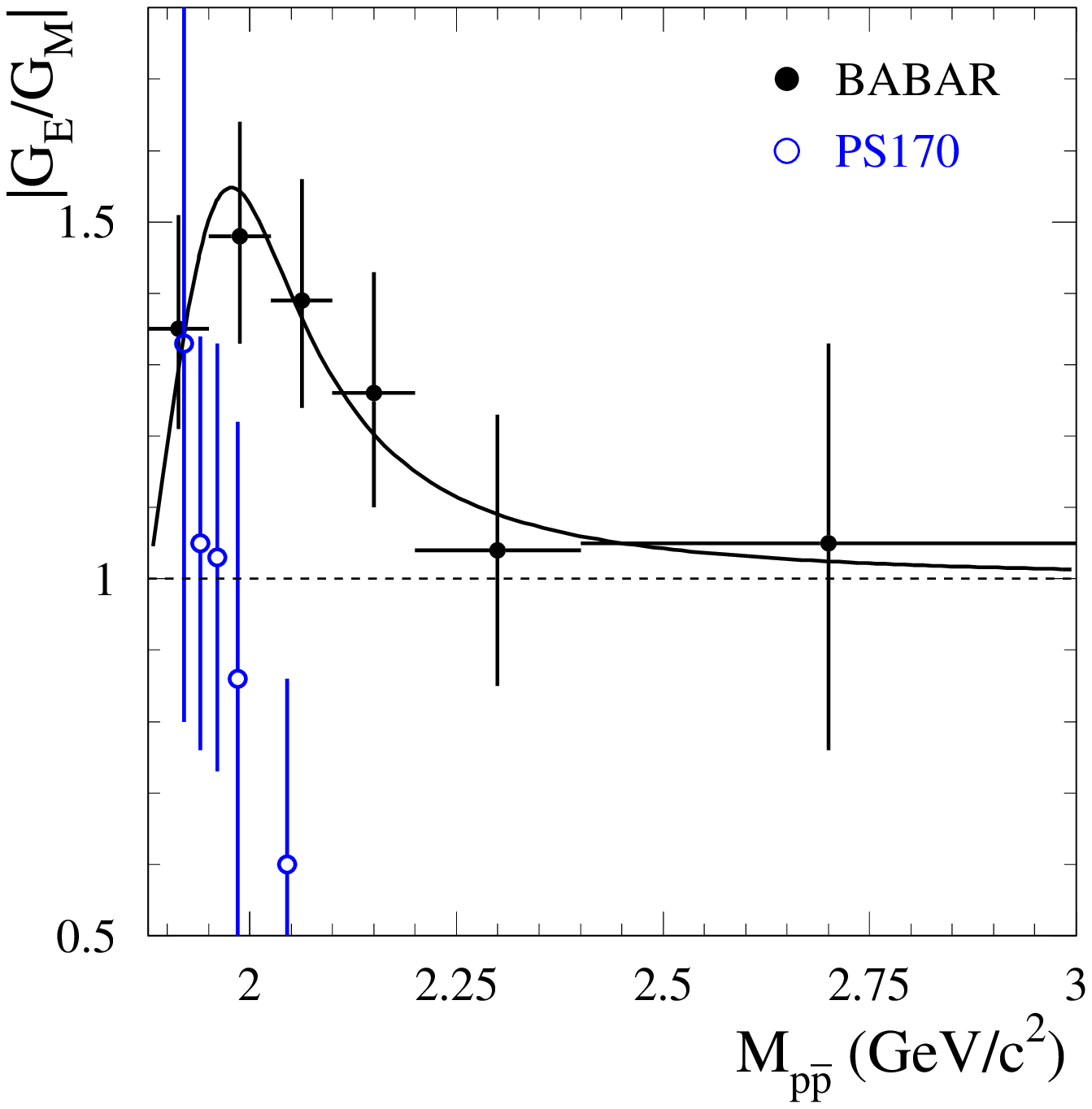}
\caption{ The measured $|G_E/G_M|$ mass dependence.
Filled circles depict \babar\ data. Open circles show
PS170 data~\cite{LEAR}. The curve is the the result of the fit
described in the text.
\label{fig10}}
\end{figure}

The fits of the histograms to the angular
distributions are shown in Fig.~\ref{fig8}; the values of
$|G_E/G_M|$ obtained are listed in Table~\ref{ang_tab} and shown in
Fig.~\ref{fig10}. The curve in Fig.~\ref{fig10}
[$1+ax/(1+bx^3)$, where $x=M_{p\bar{p}}-2m_p$ GeV/$c^2$] is used
in the iteration procedure to calculate the weight.
The quoted errors on $|G_E/G_M|$ are statistical and systematic,
respectively.
The dominant contribution to the systematic error is due to
the uncertainty in the $p\bar{p}\pi^0$ background. 

The only previous measurement of the $|G_E/G_M|$ ratio was performed
in the PS170 experiment~\cite{LEAR}. The ratio was measured at five points
between 1.92 GeV/$c^2$ and 2.04 GeV/$c^2$ with an accuracy of 
30--40\% (see Fig.~\ref{fig10}). 
For all points it was found to be consistent with unity. 
The average of the PS170 measurements evaluated under the assumption
that the errors are purely statistical is $0.90\pm0.14$.
The \babar\ results are significantly larger
for $M_{p\bar{p}} < 2.1$ GeV/$c^2$, and extend the measurements up to 
3 GeV/$c^2$.
\vspace{5mm}\\

In addition, we search for an asymmetry in the proton angular distribution.
The lowest-order one-photon mechanism for proton-antiproton production
predicts a symmetric angular distribution. An asymmetry arises
from higher-order contributions, in particular, from two-photon exchange.
Two-photon exchange is discussed (see, for example, Ref.~\cite{TPEep})
as a possible source of the difference observed in $ep$ scattering
between the $G_E/G_M$ measurements obtained with two different
experimental techniques,
namely the Rosenbluth method~\cite{Rosenbluth}, which uses the analysis of angular 
distributions,
and the polarization method~\cite{polar1,polar2,polar3}, which is based on the measurement of the
ratio of the transverse and longitudinal polarization of the recoil
proton.

A search for an asymmetry using previous \babar\ $e^+e^-\to p\bar{p}\gamma$
data~\cite{babarpp} is described in Ref.~\cite{TPEee}.
No asymmetry was observed within the statistical error of 2\%.
It should be noted that 
the authors of Ref.~\cite{TPEee} did not take into account the angular asymmetry
of the detection efficiency, which is seen in Fig.~\ref{fig9} and in a
similar plot in Ref.~\cite{babarpp}.
\begin{figure}
\includegraphics[width=.4\textwidth]{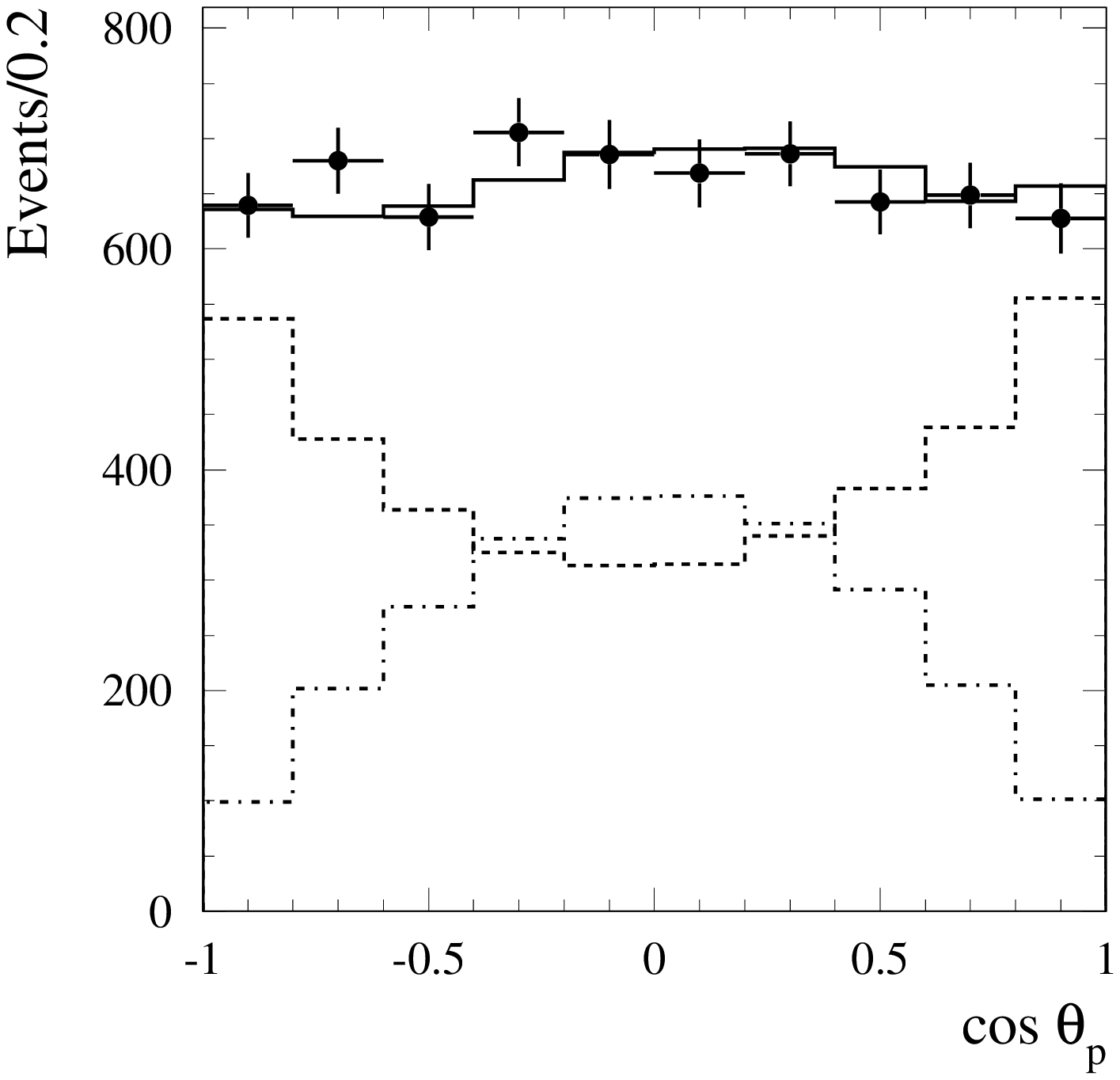}
\caption{The $\cos{\theta_p}$ distribution for
the mass region from threshold to 3 GeV/$c^2$.
The points with error bars show the data distribution after
background subtraction;
the solid histogram is the fit result. The dashed and
dot-dashed histograms show the contributions
of the terms corresponding to the magnetic and electric
form factors, respectively.
\label{fig11}}
\end{figure}
\begin{figure}
\includegraphics[width=.4\textwidth]{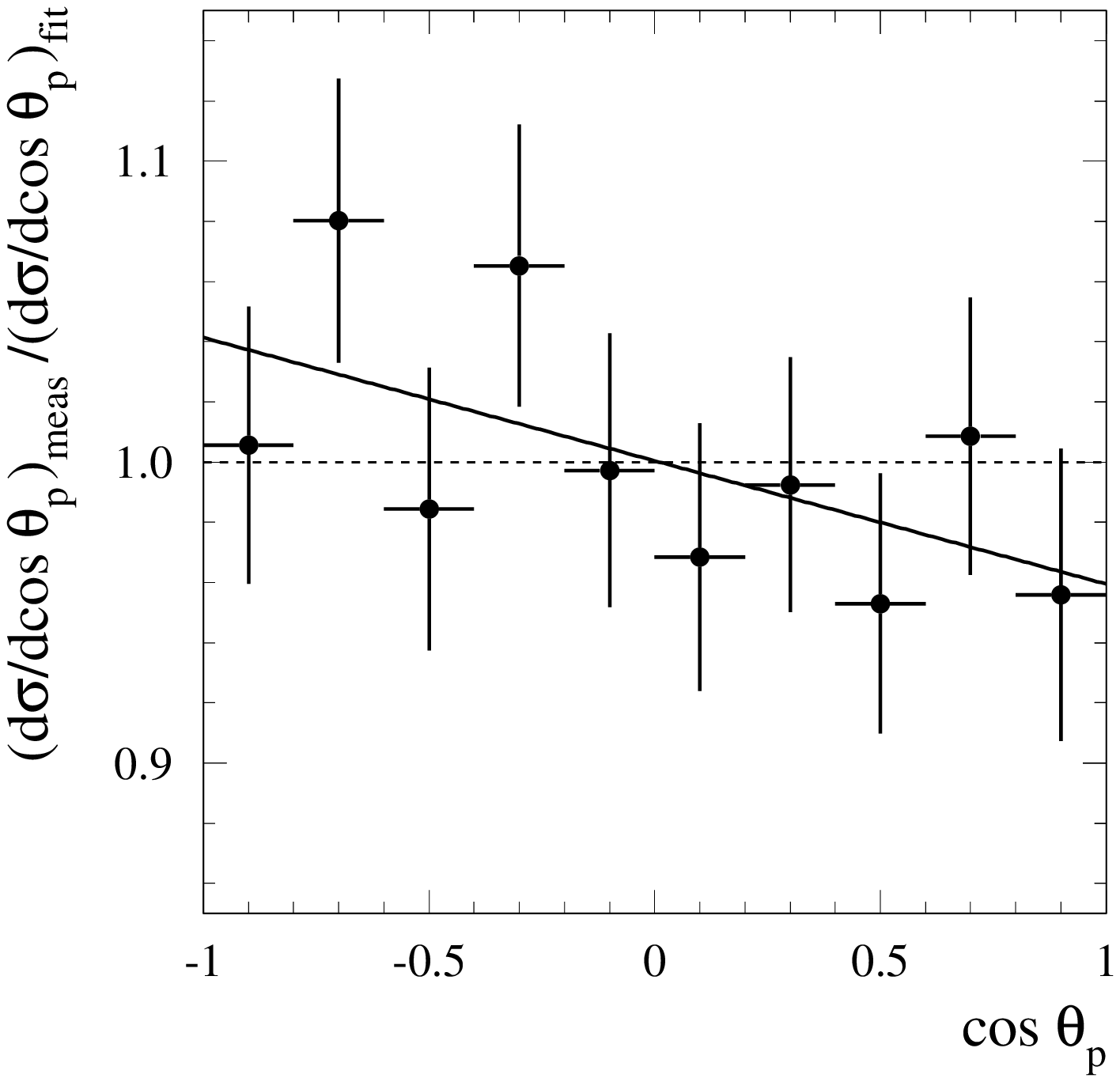}
\caption{The ratio of the data distribution from Fig.~\ref{fig11} to
the fitted simulated distribution. The line shows the result of the fit
of a linear function to the data points.
\label{fig12}}
\end{figure}
To measure the asymmetry we use the data with $p\bar{p}$ mass less than 3
GeV/$c^2$. The $\cos{\theta_p}$ distribution is fitted as described above,
and the result is shown in Fig.~\ref{fig11}. Since the MC
simulation uses a model with one-photon exchange, the asymmetry in the
fitted histogram is due to asymmetry in the detection efficiency.
To remove detector effects we take the ratio of the data distribution to
the fitted simulated distribution. This ratio is shown in Fig.~\ref{fig12}. 
A fit of a linear function to the data yields a slope parameter value
$-0.041\pm0.026\pm0.005$.
The systematic error on the slope is estimated conservatively as the
maximum slope given by an efficiency correction.
The correction for the data-MC simulation difference in
antiproton nuclear interactions (see Sec.~\ref{sdetef}) is found to yield
the largest angular variation. 

We then calculate the integral asymmetry
\begin{eqnarray}
A_{\cos{\theta_p}}&=&\frac{\sigma(\cos{\theta_p}>0)-\sigma(\cos{\theta_p}<0)}
{\sigma(\cos{\theta_p}>0)+\sigma(\cos{\theta_p}<0)}\nonumber \\
&=&-0.025\pm0.014\pm0.003,
\end{eqnarray}
where $\sigma(\cos{\theta_p}>0)$ and $\sigma(\cos{\theta_p}<0)$ are
the cross sections for $e^+e^-\to p\bar{p}\gamma$ events with
$M_{p\bar{p}} < 3$ GeV/$c^2$ integrated over the angular regions with
$\cos{\theta_p}>0$ and $\cos{\theta_p}<0$, respectively. The fitted slope value
and the integral asymmetry 
are consistent with zero.
The value of the asymmetry extracted from experiment depends on 
the selection criteria used, in particular, on the effective energy limit
for an extra photon emitted from the initial or final state. In our analysis, this
limit is determined by the condition $\chi^2_p<30$, and is about 100 MeV.

\section{Detection efficiency}\label{sdetef}
\begin{figure}
\includegraphics[width=.48\textwidth]{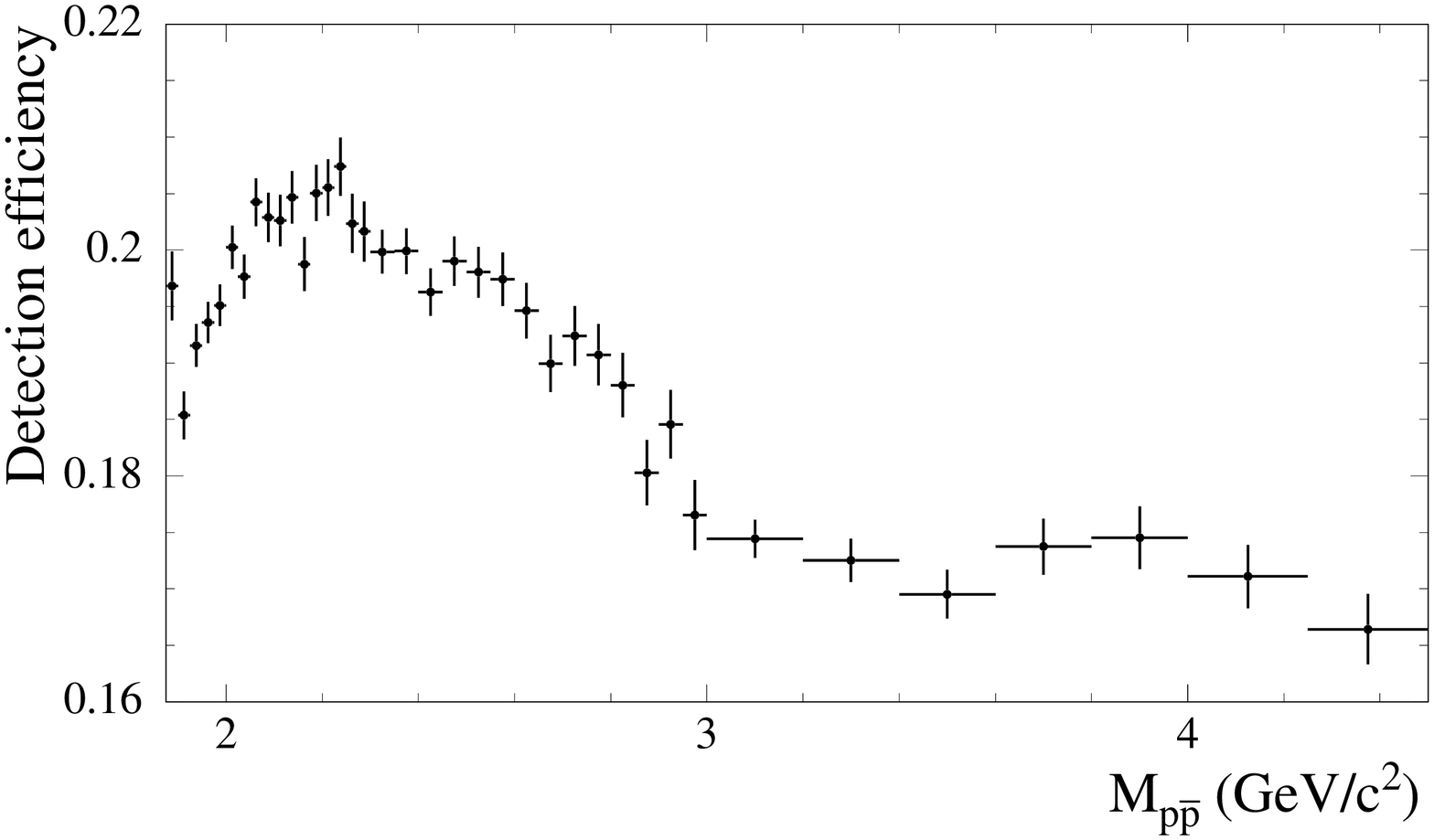}
\caption{The ${p\bar{p}}$ mass dependence of the detection
efficiency obtained from MC simulation.
\label{fig13}}
\end{figure}

The detection efficiency, which is determined using Monte Carlo 
simulation, is the ratio of true ${p\bar{p}}$ mass distributions
computed after and before applying the selection criteria. Since the
$e^+e^-\to p\bar{p} \gamma$ differential cross section depends
on two form factors, the detection efficiency cannot be determined in
a model-independent way. We use
a model with the $|G_E/G_M|$ ratio obtained from the fits to
the experimental angular distributions (curve in Fig.~\ref{fig10})
for $M_{p\bar{p}}<3$~GeV/$c^2$,  and with $|G_E/G_M|=1$ for higher masses.
The detection efficiency obtained by using this model is shown in
Fig.~\ref{fig13}.
The error in the detection efficiency due to the model is determined from the 
uncertainty in the $|G_E/G_M|$ ratio:
for $M_{p\bar{p}}<3$~GeV/$c^2$ the variation of the ratio within its 
experimental uncertainties leads to a 1\% change in the detection efficiency.
This is taken as the model-related uncertainty. For the mass region above
3~GeV/$c^2$, where the $|G_E/G_M|$ ratio is not measured, the
model uncertainty is taken as the maximum difference between
the detection efficiencies corresponding to setting $G_E=0$ or 
setting $G_M=0$, and
the efficiency calculated with the model described above. This
yields a 4\% uncertainty estimate.

The efficiency determined from MC simulation
($\varepsilon_{\rm MC}$) must be corrected to account for
data-MC simulation differences in detector response:
\begin{equation}
\varepsilon=\varepsilon_{\rm MC}\prod (1+\delta_i),
\label{eq_eff_cor}
\end{equation}
where the $\delta_i$ are efficiency corrections
for each of several effects. These corrections are
discussed in detail below and summarized in Table~\ref{tab_ef_cor}.

Inaccuracies in the simulation of angular and momentum resolution
and radiative corrections may account for data-MC
differences in the fraction of events rejected by the requirement 
$\chi^2_p<30$.
The efficiency correction for this effect is estimated by comparing
data and simulated $\chi^2$ distributions for the $e^+e^-\to \mu^+\mu^-\gamma$
process, which has kinematics similar to the process under study.
An exclusive $e^+e^-\to \mu^+\mu^-\gamma$ sample is  selected by requiring that
both charged tracks be identified as muons. 
The ratio of the number of
selected muon events with $\chi^2_\mu>30$ and $\chi^2_\mu<30$
varies from 0.30 to 0.37 in the
$M_{p \bar p}$ range from threshold to 4.5~GeV/$c^2$.
To characterize data-MC simulation differences in the $\chi^2$ distribution,
a double ratio ($\kappa$) is calculated as the ratio 
of $N(\chi^2_\mu>30)/N(\chi^2_\mu<30)$ 
obtained from data to the same quantity obtained from MC simulation.
The value of the double ratio varies from 1.02 to 1.06
in the $M_{p \bar p}$ range from threshold to 4.5~GeV/$c^2$.
The efficiency correction $\delta_i$ (with $i=1$) for the $\chi^2_p$ cut is 
calculated as
\begin{equation}
\delta_1=\frac{N(\chi^2_p<30)+N(\chi^2_p>30)}{N(\chi^2_p<30)+\kappa
N(\chi^2_p>30)}-1,  
\end{equation}
where $N(\chi^2_p<30)$ and $N(\chi^2_p>30)$ are the numbers of simulated
$p\bar{p}\gamma$ events with
$\chi^2_p<30$ and $\chi^2_p>30$, respectively.
The values of the efficiency correction $\delta_1$
for different $p\bar{p}$ invariant-mass values are listed in 
Table~\ref{tab_ef_cor}.

The effect of the $\chi^2_K>30$ requirement is studied using 
$e^+e^-\to J/\psi\gamma \to p\bar{p}\gamma$ events. The 
$J/\psi$ yield is determined using the sideband subtraction method.
The event losses are found to be
$(1.5\pm 0.4)\%$ in data and $(1.2\pm 0.1)\%$ in MC simulation.
As the data and simulated values are in good agreement, there is no 
need to  introduce any efficiency correction for the $\chi^2_K>30$ requirement.
The systematic uncertainty associated with this criterion is 0.4\%. 

      Another possible source of data-MC simulation differences is 
is due to unreconstructed tracks.
Two dominant effects leading to track loss in $p\bar{p}\gamma$
events are track overlap in the DCH and nuclear interaction
of  protons and antiprotons in the material before the SVT and DCH.

\begin{figure}
\includegraphics[width=.4\textwidth]{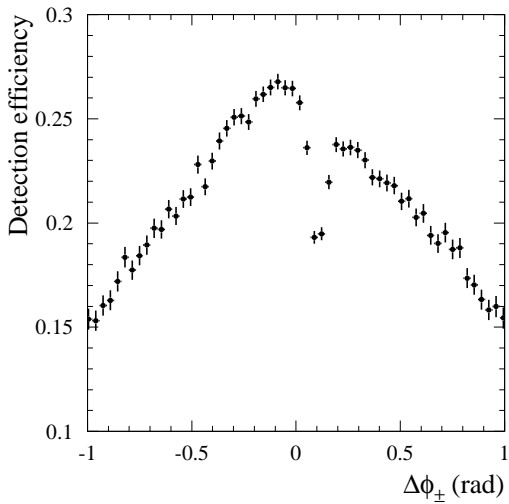}
\caption{The detection efficiency for $e^+e^-\to p\bar{p}\gamma$ events
as a function of $\Delta \varphi_{\pm}$ obtained from MC simulation.
\label{fig14}}
\end{figure}

The effect of track overlap can be observed in the distribution of 
the parameter
$\Delta \varphi_{\pm}=\varphi_+ - \varphi_-$,
where $\varphi_+$ and $\varphi_-$ are the azimuthal angles at the 
production vertex of
positive and negative tracks, respectively. The detection efficiency for
simulated $e^+e^-\to p\bar{p}\gamma$ events as a function of 
$\Delta \varphi_{\pm}$ is shown in Fig.~\ref{fig14}. 

The $z$-component of the \babar\ magnetic field is in the direction of the 
positive $z$-axis, so that in the $x-y$ plane viewed from positive $z$
positively charged tracks experience 
clockwise bending, while negatively charged tracks are bent counter-clockwise.
As a result, events with $\Delta \varphi_{\pm}>0$ have a ``fishtail'' two-track
configuration in which the tracks tend to overlap initially.
This results in the dip in efficiency which is clearly seen at 
$\Delta \varphi_{\pm}\sim 0.1$ rad. The
ratio of the number of events with $\Delta \varphi_{\pm}>0$ to that with
$\Delta \varphi_{\pm}<0$ can be used to estimate the efficiency loss
due to track overlap. This efficiency loss reaches about 10\% near
$p\bar{p}$ threshold and decreases to a negligible level for $M_{p\bar{p}}$
above 2.4 GeV/$c^2$. The effect is reproduced reasonably well by
the MC simulation; data-MC simulation differences in the efficiency loss
averaged over the mass region of maximum inefficiency, 
$M_{p\bar{p}}<2.3$ GeV/$c^2$, is about $(1.2\pm 1.3)\%$. 
We introduce no correction for this difference. For the mass  
region $M_{p\bar{p}}<2.3$ GeV/$c^2$, where the effect is large,
a systematic uncertainty of 1.5\% is assigned to the measured cross section.

Simulation shows that nuclear interaction leads to the loss of 
approximately 6\% of the $e^+e^-\to p\bar{p}\gamma$  events. 
For data-MC simulation comparison,  a specially
selected event sample with $\Lambda(\bar{\Lambda})$ decaying
into $p(\bar{p})\pi$ is used. The $\Lambda$ candidates are selected by imposing
requirements on the $p\pi$ invariant mass and $\Lambda$ flight distance.
The amount of material before the SVT
(1.5\% of a nuclear interaction length) is comparable
to the amount of material between the SVT and the DCH 
(1.4\% of a nuclear interaction length).
The probability of track losses between the SVT and the DCH
is measured by using the $\Lambda(\bar{\Lambda})$ sample.
The data and simulation probabilities are found to be in
agreement for protons. 
A substantial difference is observed
for antiprotons, which is consistent with a
large (a factor of $2.6\pm1.0$) overestimation of the antiproton annihilation 
cross section in simulation. 
This difference in the antiproton
annihilation cross section in data and simulation leads to a correction of
about $(1.0\pm0.4)\%$ to the detection efficiency for $p\bar{p}\gamma$ events. 

We also incorporate a systematic uncertainty due to data-MC simulation 
differences in track reconstruction, which is estimated to be 0.24\% per track.

The data-MC simulation difference in particle identification 
is studied using events with a $J/\psi\to p\bar{p}$ decay.
Due to the narrow $J/\psi$ width
and hence low background, the  number of $J/\psi\to p\bar{p}$ decays
may  be determined using selections with either one or two identified
protons.    The background from non-$J/\psi$ events is
subtracted using sidebands.
The $p / \bar{p}$ identification probabilities
are determined as functions of the $p / \bar{p}$ momenta
by calculating the ratio of the number of events
with both the proton and the antiproton identified to the number of
events with only one identified proton or antiproton.
The ratio of data-MC identification probabilities is used to reweight
selected simulated events and calculate efficiency corrections.
The correction is about $-(1.9\pm 2.0)\%$ 
and is practically independent of $p\bar{p}$ mass.
The error in the correction is determined 
from the statistical uncertainty on the number of selected $J/\psi$ events.

An additional correction must be applied to the photon detection efficiency. 
There are two main sources of this correction. The first is due to data-MC
simulation differences in the probability of photon conversion in the detector
material before the DCH, and the second results from the effect of dead
calorimeter channels.
A sample of $e^+e^-\to \mu^+\mu^-\gamma$ events is 
used to determine the calorimeter photon inefficiency in data. 
Events with exactly two oppositely charged particle 
tracks identified as muons are selected, and a one-constraint kinematic fit is 
performed, in which the recoil mass against the muon pair is constrained
to be zero. A tight requirement on the 
$\chi^2$ of the kinematic fit selects events with only one photon in the final 
state. The photon  direction is determined from the fit, and the detection
inefficiency is calculated as the ratio of the number of events not
satisfying the  $E_\gamma^\ast > 3$~GeV criterion,
to the total number of selected $\mu^+\mu^-\gamma$ events.
The photon inefficiency obtained is 3.3\%, to be compared to the 2\% 
inefficiency from the $e^+e^-\to \mu^+\mu^-\gamma$
simulation. 
The data-MC simulation difference in the probability of photon conversion
is also studied using $e^+e^-\to \mu^+\mu^-\gamma$ events.
In addition to two identified muons, we require that an event contain
a converted-photon candidate, i.e., a pair of oppositely charged tracks 
with $e^+ e^-$ invariant mass close to zero, momentum directed along 
the expected photon direction, and forming a secondary vertex well-separated
from the interaction region. The observed data-MC difference
in the probability of photon conversion is $-(0.41\pm0.01)\%$. 
The data-MC differencies in the calorimeter inefficiency for photons and 
the probability of photon conversion are determined as functions of the 
photon polar angle, and used to reweight the simulated events and calculate 
efficiency corrections.
The total correction due to data-MC simulation differences in the photon
detection inefficiency is found to vary from of $-(1.9\pm0.1)\%$ near 
the $p\bar{p}$ threshold to $-(1.7\pm0.1)\%$ at 3 GeV/$c^2$ and 
higher masses.

The quality of the simulation of trigger efficiency is also studied. The
overlap of the samples of events satisfying different
trigger criteria, and the independence of these triggers, are used to measure
trigger efficiency. The difference in trigger efficiency between data and 
MC simulation decreases from $-(0.65\pm0.20)\%$ near $p\bar{p}$ threshold to 
$-(0.13\pm0.10)\%$ for $p\bar{p}$ masses above 2.2 GeV/$c^2$.
An additional systematic uncertainty of about 0.5\% is introduced to
take into account the possibly imperfect simulation of inefficiency
of the offline filters that provide background
suppression before full event reconstruction.

All efficiency corrections are summarized in Table~\ref{tab_ef_cor}, and
the corrected detection efficiency values are listed in
Table~\ref{sumtab}. The uncertainty in  detection efficiency
includes simulation statistical error, model uncertainty,
and the uncertainty on the efficiency correction.
\begin{table}
\caption{The values of the different efficiency corrections $\delta_i$
for $p \bar{p}$ invariant mass 1.9, 3.0, and 4.5~GeV/$c^2$.
\label{tab_ef_cor}}
\begin{ruledtabular}
\begin{tabular}{lccc}
  effect              &$\delta_i(1.9),\%$&$\delta_i(3),\%$&$\delta_i(4.5),\%$\\[0.3ex]
  \hline
  \\[-2.1ex]
  $\chi^2_p<30$        & $-0.5\pm0.1$  & $-0.9\pm0.1$ &$-1.5\pm0.2$\\
 $\chi^2_K>30$        & $ 0.0\pm0.4$  & $ 0.0\pm0.4$ &$ 0.0\pm0.4$\\
 track overlap        & $ 0.0\pm1.5$  &  --          & --         \\
 nuclear interaction  & $ 0.8\pm0.4$  & $ 1.1\pm0.4$ &$ 1.0\pm0.4$\\
 track reconstruction & $ 0.0\pm0.5$  & $ 0.0\pm0.5$ &$ 0.0\pm0.5$\\
 PID                  & $-1.9\pm2.0$  & $-1.9\pm2.0$ &$-1.9\pm2.0$\\
 photon inefficiency  & $-1.9\pm0.1$  & $-1.7\pm0.1$ &$-1.7\pm0.1$\\
 trigger and filters  & $-0.7\pm0.6$  & $-0.1\pm0.5$ &$-0.1\pm0.5$     \\[0.3ex]
 \hline
 \\[-2.1ex]
  total                & $-4.2\pm2.6$  & $-3.5\pm2.2$ & $-4.2\pm2.2$\\
 \end{tabular}
\end{ruledtabular}
 \end{table}
\section{\boldmath The $e^+e^-\to p\bar{p}$
cross section and the proton form factor\label{xsec}}
The cross section for $e^+e^-\to p\bar{p}$ is calculated from
the $p\bar{p}$ mass spectrum using the expression
\begin{equation}
\sigma_{p\bar{p}}(M_{p\bar{p}})=\frac{(dN/dM_{p\bar{p}})_{corr}}{\varepsilon\, R\,
dL/dM_{p\bar{p}}},
\end{equation}
where $(dN/dM_{p\bar{p}})_{corr}$ is the mass spectrum corrected for
resolution effects,
${dL}/{dM_{p\bar{p}}}$ is the ISR differential
luminosity, $\varepsilon(M_{p\bar{p}})$ is the detection efficiency as a function of mass,
and $R$ is a radiative correction factor accounting for the Born mass
spectrum distortion due to emission of extra photons by the initial
       electron and positron. The ISR luminosity is calculated
using the total integrated luminosity $L$ and the integral over 
$\cos{\theta_\gamma^\ast}$ of the probability density
function for ISR photon emission (Eq.~(\ref{eq2})):
\begin{equation}
\frac{dL}{dM_{p\bar{p}}}=\frac{\alpha}{\pi x}\left(
(2-2x+x^2)\log\frac{1+B}{1-B}-x^2 C\right)\frac{2M_{p\bar{p}}}{s}\,L.
\label{ISRlum}
\end{equation}
Here $B=\cos{\theta_0^\ast}$, and $\theta_0^\ast$ determines the range of 
polar angles for the ISR photon in the $e^+e^-$ c.m.~frame:
$\theta_0^\ast<\theta_\gamma^\ast<180^\circ-\theta_0^\ast$. 
In our case $\theta_0^\ast=20^\circ$,
since we determine detector efficiency using simulation with
$20^\circ<\theta_\gamma^\ast<160^\circ$. The values of ISR luminosity
integrated over the $M_{p\bar{p}}$ intervals are listed in Table~\ref{sumtab}.

The radiative correction factor $R$ is determined from MC
simulation at the generator level, with no detector simulation.
The  $p\bar{p}$ mass spectrum is generated using only
the pure Born amplitude for the process $e^+ e^- \to p\bar{p}\gamma $,
and then using a model with higher-order
radiative corrections included by means of the structure function
method~\cite{strfun}.
The radiative correction factor, evaluated as the ratio of the second 
spectrum to the first, varies from 1.001 at $p\bar{p}$ threshold to 
1.02 at $M_{p\bar{p}}=4.5$~GeV/$c^2$.

The value of $R$ depends on the requirement
on the invariant mass of the $p\bar{p}\gamma$ system.
The value of $R$ obtained in our case corresponds to the requirement
$M_{p\bar{p}\gamma} > 8$~GeV/$c^2$ imposed in the simulation.
The theoretical uncertainty on the radiative correction calculation
by the structure function method does not exceed 1\%~\cite{strfun}.
The calculated radiative correction factor does not take into account 
vacuum polarization; the contribution of the latter is included in the
measured cross section.

\begin{figure}
\includegraphics[width=.45\textwidth]{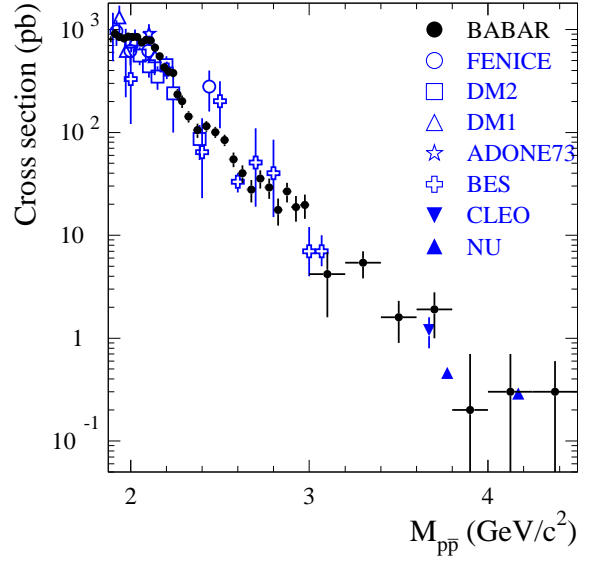}
\caption{The $e^+e^-\to p\bar{p}$ cross section measured in this analysis
and in other
$e^+e^-$ experiments: FENICE\cite{FENICE}, DM2\cite{DM2}, DM1\cite{DM1},
ADONE73\cite{ADONE73}, BES\cite{BES}, CLEO\cite{CLEO}, NU\cite{CLEO2012}.
The contributions of $J/\psi\to p\bar{p}$ and $\psi (2S)\to p\bar{p}$
decays to the \babar\ measurement have been subtracted.} 
\label{fig15}
\end{figure}
\begin{figure}
\includegraphics[width=.45\textwidth]{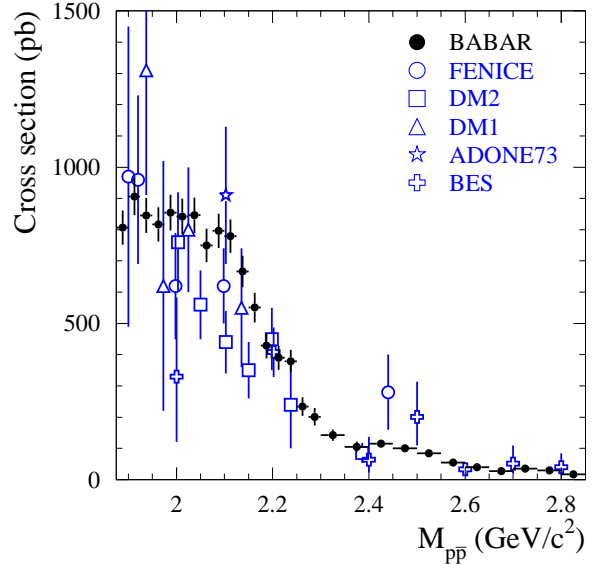}
\caption{The $e^+e^-\to p\bar{p}$ cross section near threshold 
measured in this analysis and in other
$e^+e^-$ experiments: FENICE\cite{FENICE}, DM2\cite{DM2}, DM1\cite{DM1},
ADONE73\cite{ADONE73}, BES\cite{BES}.
\label{fig16}}
 \end{figure}

 \begin{table*}
\caption{ The $p\bar{p}$ invariant-mass interval ($M_{p\bar{p}}$),
number of selected events ($N$) after  background subtraction and mass migration,
detection efficiency ($\varepsilon$), ISR luminosity ($L$), 
measured cross section ($\sigma_{p\bar{p}}$), and $|F_p|$, the effective form 
factor for $e^+e^-\to p\bar{p}$.
The contributions from $J/\psi\to p\bar{p}$ and $\psi (2S)\to p\bar{p}$
decays have been subtracted. The quoted uncertainties on $N$ and $\sigma$ are 
statistical and systematic, respectively.
For the form factor, the combined uncertainty is listed.
\label{sumtab}} 
\begin{ruledtabular}
\begin{tabular}{cccccc}
$M_{p\bar{p}}$ (GeV/$c^2$)&$N$&$\varepsilon$&$L$ (pb$^{-1}$)&$\sigma_{p\bar{p}}$ (pb)&$|F_p|$\\[0.3ex]
\hline
\\[-2.1ex]
1.877--1.900&$351\pm 20\pm 4$&$0.189\pm0.006$& 2.33&$806\pm 46\pm 30$&$0.424\pm0.014$\\
1.900--1.925&$403\pm 22\pm 4$&$0.178\pm0.006$& 2.52&$906\pm 50\pm 32$&$0.355\pm0.012$\\
1.925--1.950&$394\pm 22\pm 5$&$0.184\pm0.006$& 2.56&$845\pm 47\pm 31$&$0.309\pm0.010$\\
1.950--1.975&$390\pm 22\pm 5$&$0.186\pm0.006$& 2.60&$817\pm 46\pm 30$&$0.286\pm0.010$\\
1.975--2.000&$418\pm 24\pm 5$&$0.187\pm0.006$& 2.63&$854\pm 48\pm 31$&$0.281\pm0.009$\\
2.000--2.025&$429\pm 24\pm 5$&$0.192\pm0.006$& 2.67&$842\pm 48\pm 30$&$0.271\pm0.009$\\
2.025--2.050&$433\pm 24\pm 6$&$0.191\pm0.006$& 2.71&$846\pm 48\pm 31$&$0.266\pm0.009$\\
2.050--2.075&$402\pm 24\pm 7$&$0.197\pm0.006$& 2.75&$750\pm 45\pm 28$&$0.247\pm0.009$\\
2.075--2.100&$430\pm 25\pm 6$&$0.196\pm0.006$& 2.79&$796\pm 46\pm 29$&$0.252\pm0.009$\\
2.100--2.125&$426\pm 25\pm 6$&$0.195\pm0.006$& 2.83&$779\pm 45\pm 29$&$0.247\pm0.008$\\
2.125--2.150&$373\pm 24\pm 8$&$0.197\pm0.006$& 2.86&$666\pm 43\pm 27$&$0.227\pm0.009$\\
2.150--2.175&$304\pm 22\pm 8$&$0.192\pm0.006$& 2.90&$551\pm 41\pm 24$&$0.206\pm0.009$\\
2.175--2.200&$247\pm 20\pm 8$&$0.198\pm0.006$& 2.94&$429\pm 35\pm 20$&$0.182\pm0.009$\\
2.200--2.225&$228\pm 20\pm 8$&$0.198\pm0.006$& 2.98&$390\pm 33\pm 19$&$0.173\pm0.008$\\
2.225--2.250&$227\pm 19\pm 6$&$0.200\pm0.006$& 3.02&$379\pm 32\pm 16$&$0.171\pm0.008$\\
2.250--2.275&$139\pm 16\pm 6$&$0.195\pm0.006$& 3.06&$234\pm 27\pm 13$&$0.134\pm0.009$\\
2.275--2.300&$120\pm 15\pm 6$&$0.195\pm0.006$& 3.10&$201\pm 25\pm 12$&$0.125\pm0.009$\\
2.300--2.350&$173\pm 17\pm13$&$0.193\pm0.005$& 6.32&$143\pm 14\pm 12$&$0.106\pm0.007$\\
2.350--2.400&$130\pm 15\pm13$&$0.193\pm0.005$& 6.48&$105\pm 12\pm 11$&$0.091\pm0.007$\\
2.400--2.450&$143\pm 15\pm 5$&$0.190\pm0.005$& 6.64&$115\pm 12\pm  6$&$0.096\pm0.006$\\
2.450--2.500&$131\pm 15\pm 5$&$0.192\pm0.005$& 6.80&$101\pm 11\pm  5$&$0.091\pm0.006$\\
2.500--2.550&$111\pm 13\pm 4$&$0.191\pm0.005$& 6.97&$ 84\pm 10\pm  4$&$0.084\pm0.005$\\
2.550--2.600&$ 74\pm 11\pm 4$&$0.191\pm0.005$& 7.14&$ 55\pm  8\pm  3$&$0.069\pm0.006$\\
2.600--2.650&$ 55\pm 10\pm 3$&$0.188\pm0.005$& 7.31&$ 40\pm  8\pm  3$&$0.060\pm0.006$\\
2.650--2.700&$ 38\pm  9\pm 3$&$0.183\pm0.005$& 7.48&$ 28\pm  6\pm  3$&$0.050\pm0.006$\\
2.700--2.750&$ 50\pm  9\pm 3$&$0.186\pm0.005$& 7.66&$ 36\pm  7\pm  3$&$0.058\pm0.006$\\
2.750--2.800&$ 42\pm  9\pm 3$&$0.184\pm0.005$& 7.84&$ 29\pm  6\pm  3$&$0.053\pm0.006$\\
2.800--2.850&$ 25\pm  7\pm 2$&$0.181\pm0.005$& 8.01&$ 18\pm  5\pm  1$&$0.042\pm0.006$\\
2.850--2.900&$ 38\pm  8\pm 2$&$0.174\pm0.005$& 8.20&$ 27\pm  6\pm  2$&$0.052\pm0.006$\\
2.900--2.950&$ 28\pm  7\pm 2$&$0.178\pm0.005$& 8.38&$ 19\pm  5\pm  2$&$0.044\pm0.006$\\
2.950--3.000&$ 29\pm  7\pm 2$&$0.170\pm0.005$& 8.57&$ 20\pm  5\pm  2$&$0.046\pm0.006$\\
3.000--3.200&$ 25\pm 12\pm 9$&$0.168\pm0.008$&36.19&$4.2\pm2.0\pm1.6$&$0.022\pm0.007$\\
3.200--3.400&$ 36\pm  8\pm 7$&$0.166\pm0.008$&39.40&$5.4\pm1.2\pm1.1$&$0.027\pm0.004$\\
3.400--3.600&$ 11\pm  4\pm 2$&$0.163\pm0.008$&42.81&$1.6\pm0.6\pm0.3$&$0.015\pm0.003$\\
3.600--3.800&$ 15\pm  6\pm 2$&$0.167\pm0.008$&46.44&$1.9\pm0.8\pm0.3$&$0.018\pm0.004$\\
3.800--4.000&$  1\pm  3\pm 2$&$0.168\pm0.008$&50.33&$0.2\pm0.4\pm0.2$&$0.005\pm0.005$\\
4.000--4.250&$  4\pm  3\pm 2$&$0.164\pm0.008$&68.83&$0.3\pm0.3\pm0.2$&$0.008_{-0.008}^{+0.004}$\\
\\[-2.1ex]
4.250--4.500&$  3\pm  4\pm 2$&$0.160\pm0.008$&76.00&$0.3\pm0.3\pm0.2$&$0.008_{-0.008}^{+0.004}$\\
\end{tabular}
\end{ruledtabular}
\end{table*}

The resolution-corrected mass spectrum is obtained by unfolding
the mass resolution from the measured mass spectrum. 
Using MC simulation, a migration matrix, $A$, is obtained, 
which represents the probability that an event with  true mass 
($M_{p\bar{p}}^{true}$) in mass interval $j$ is reconstructed in interval $i$:
\begin{equation}
\left( \frac{dN}{dM_{p\bar{p}}}\right)^{rec}_i=
\sum_j A_{ij}\left(\frac{dN}{dM_{p\bar{p}}}\right)^{true}_j.
\end{equation}
The mass resolution changes from 1.5 MeV/$c^2$ near threshold to 
12 MeV/$c^2$ at $M_{p\bar{p}}=3$ GeV/$c^2$ and 22 MeV/$c^2$  at 4.5 GeV/$c^2$.
Since the chosen mass interval width significantly exceeds the
resolution for all $p\bar{p}$ masses, the
migration matrix is nearly diagonal, with the values of
diagonal elements  $ \sim 0.9$,
and next-to-diagonal $ \sim 0.05$.
We unfold the mass spectrum by applying
the inverse of the migration matrix to the measured
spectrum. The procedure changes 
the shape of the mass distribution insignificantly,
but increases the uncertainties (by $\approx $20\%)
and their correlations. 

     After applying the migration matrix, the number of events in each mass 
interval is listed in Table~\ref{sumtab}. The quoted errors are statistical 
and systematic, respectively. The latter is due to the
uncertainty in background subtraction. The calculated cross section for 
$e^+e^-\to p\bar{p}$ is shown in Fig.~\ref{fig15}  and listed in 
Table~\ref{sumtab}.
For mass intervals 3--3.2~GeV/$c^2$ and 3.6--3.8~GeV/$c^2$, the nonresonant 
cross section is quoted after excluding the $J/\psi$ and $\psi(2S)$ 
contributions (see Sec.~\ref{jpsi}). The errors quoted are statistical and
systematic. The systematic uncertainty includes the uncertainty on 
the number of signal
events, detection efficiency, the total integrated luminosity
(1\%), and the radiative corrections (1\%). A comparison 
of this result with the available $e^+e^-$ data is shown in Fig.~\ref{fig15}, 
and the behavior in the near-threshold region is shown in Fig.~\ref{fig16}.

\begin{figure}
\includegraphics[width=.96\linewidth]{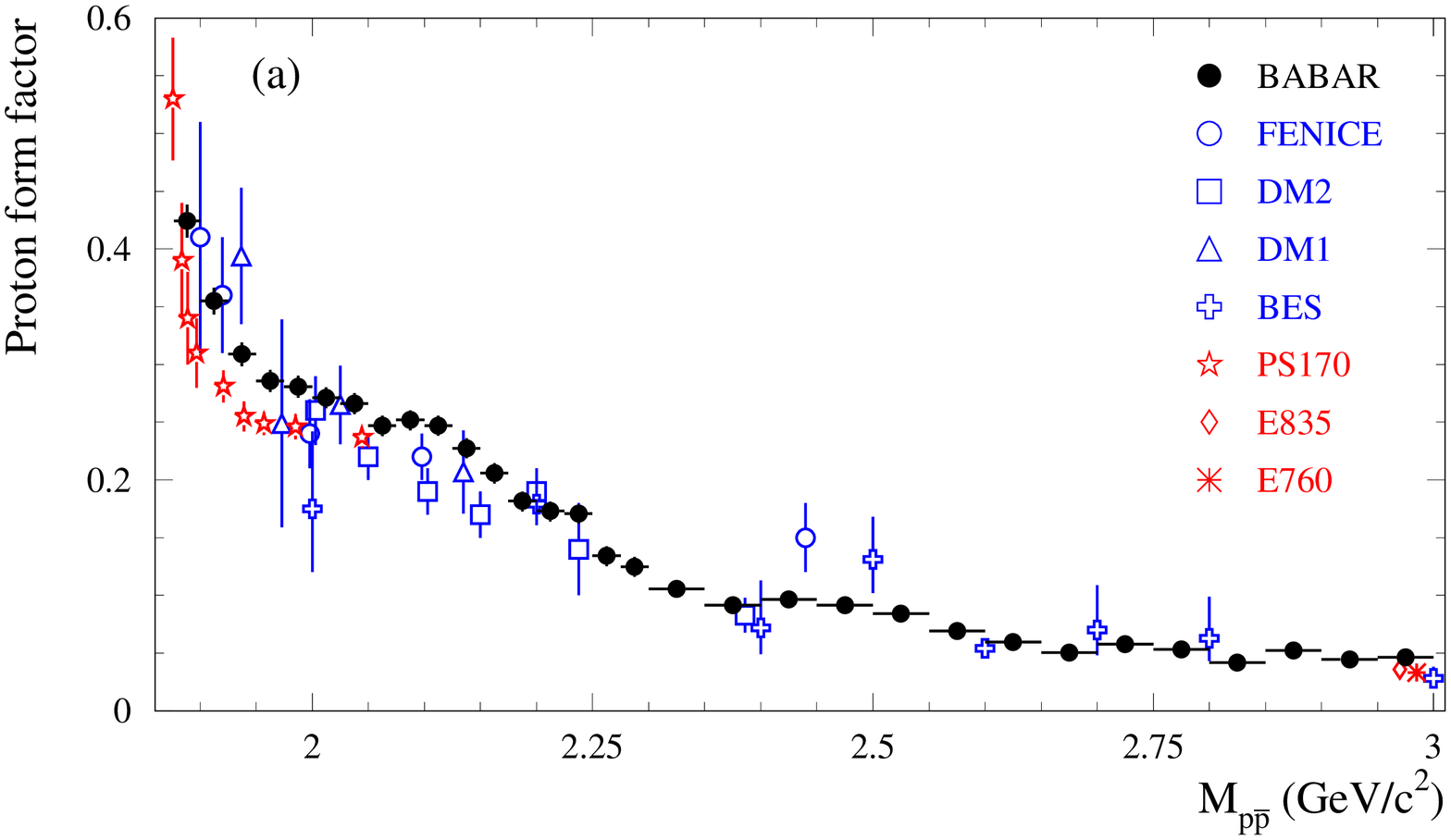}
\includegraphics[width=.96\linewidth]{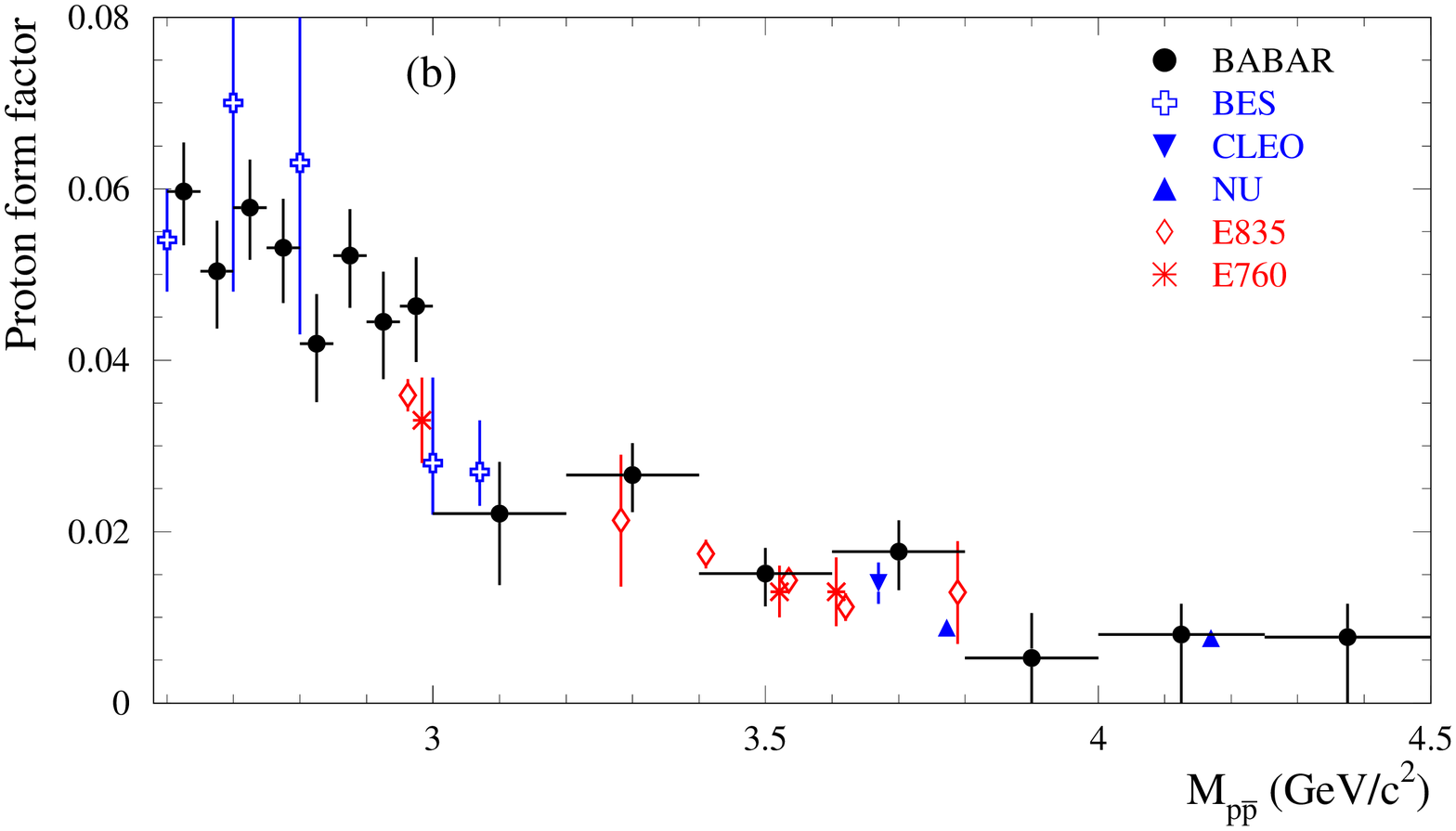}
\caption{The proton effective form factor measured in this analysis,
in other $e^+e^-$ experiments, and in $p\bar{p}$ experiments: 
FENICE\cite{FENICE}, DM2\cite{DM2},
DM1\cite{DM1}, BES\cite{BES}, CLEO\cite{CLEO}, NU\cite{CLEO2012}, PS170\cite{LEAR},
E835\cite{E835}, E760\cite{E760}: (a) for the mass interval
from $p\bar{p}$ threshold to 3.01 GeV/$c^2$, and (b) for $p\bar{p}$ masses 
from 2.58 to 4.50 GeV/$c^2$.
\label{fig17}}
\end{figure}

\begin{figure}
\includegraphics[width=.45\textwidth]{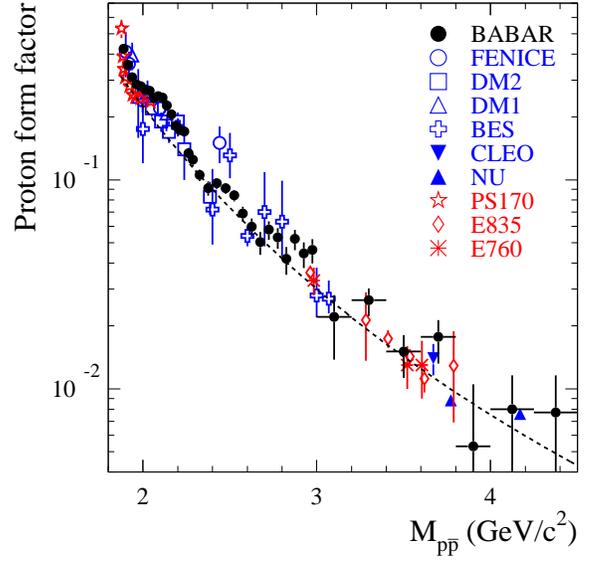}
\caption{The proton effective form factor measured in this analysis,
in other $e^+e^-$ experiments, and in $p\bar{p}$ experiments, 
shown on a logarithmic scale: 
FENICE\cite{FENICE}, DM2\cite{DM2}, DM1\cite{DM1}, BES\cite{BES}, 
CLEO\cite{CLEO}, NU\cite{CLEO2012},
PS170\cite{LEAR}, E835\cite{E835}, E760\cite{E760}.
The curve corresponds to the QCD-motivated fit described in the text.
\label{fig18}}
\end{figure}

\begin{figure}
\includegraphics[width=.45\textwidth]{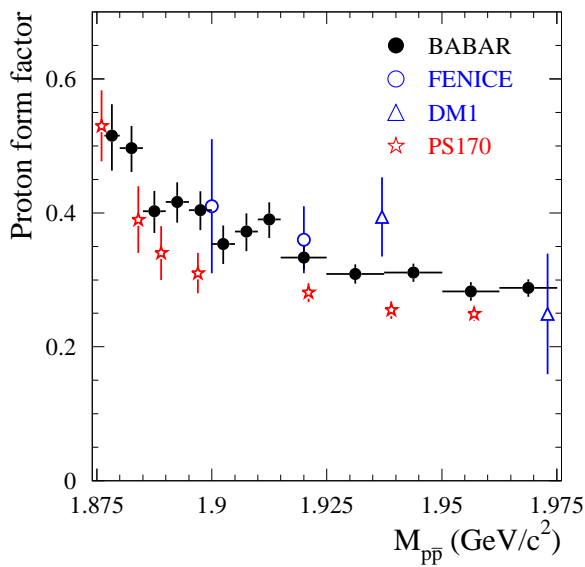}
\caption{The proton effective form factor near $p\bar{p}$ threshold
measured in this work and
in other $e^+e^-$ and $p\bar{p}$ experiments: FENICE\cite{FENICE},
DM1\cite{DM1}, PS170\cite{LEAR}.
\label{fig19}}
\end{figure}

The $e^+e^-\to p\bar{p}$ cross section is a function of two form factors,
but due to the poor determination of the  $|G_E/G_M|$ ratio, they cannot 
be extracted from the data simultaneously with reasonable accuracy.
Therefore, the effective form factor $F_p(M_{p\bar{p}})$ is introduced (Eq.~(\ref{eq4})),
which is proportional to the square root of the measured cross section.
This definition of the effective form factor permits comparison of our
measurement with measurements from other experiments, most of 
which were made under the assumption $|G_E|=|G_M|$.
The calculated effective form factor is shown in Fig.~\ref{fig17} 
(linear scale) and Fig.~\ref{fig18} (logarithmic scale),
while numerical values are listed in Table~\ref{sumtab}.
\begin{table*}
\caption{The $p\bar{p}$ invariant-mass interval ($M_{p\bar{p}}$),
number of selected events ($N$) after background subtraction and mass migration,
measured cross section ($\sigma_{p\bar{p}}$), and effective form factor 
for $e^+e^-\to p\bar{p}$ ($|F_p|$).
The quoted errors on $N$ and $\sigma_{p\bar{p}}$ are statistical and 
systematic, respectively.
For the effective form factor, the combined error is listed.
\label{tablow}}
\begin{ruledtabular}
\begin{tabular}{cccc}
$M_{p\bar{p}}$ (GeV/$c^2$)&$N$&$\sigma_{p\bar{p}}$ (pb) & $|F_p|$ \\[0.3ex]
\hline
\\[-2.1ex]
1.8765--1.8800&$  37\pm  7\pm 1$&$ 534\pm 94\pm 39$&$0.515\pm0.050$\\
1.8800--1.8850&$  80\pm 10\pm 1$&$ 826\pm106\pm 42$&$0.497\pm0.034$\\
1.8850--1.8900&$  67\pm 10\pm 1$&$ 705\pm105\pm 33$&$0.403\pm0.032$\\
1.8900--1.8950&$  79\pm 11\pm 1$&$ 886\pm121\pm 41$&$0.416\pm0.030$\\
1.8950--1.9000&$  86\pm 12\pm 1$&$ 938\pm128\pm 42$&$0.404\pm0.029$\\
1.9000--1.9050&$  70\pm 11\pm 1$&$ 785\pm123\pm 35$&$0.353\pm0.029$\\
1.9050--1.9100&$  80\pm 11\pm 1$&$ 937\pm135\pm 41$&$0.372\pm0.028$\\
1.9100--1.9150&$  98\pm 13\pm 1$&$1096\pm142\pm 46$&$0.390\pm0.027$\\
1.9150--1.9250&$ 156\pm 15\pm 2$&$ 862\pm 84\pm 32$&$0.333\pm0.017$\\
1.9250--1.9375&$ 188\pm 16\pm 3$&$ 811\pm 69\pm 31$&$0.309\pm0.014$\\
1.9375--1.9500&$ 208\pm 17\pm 3$&$ 887\pm 72\pm 33$&$0.311\pm0.014$\\
1.9500--1.9625&$ 181\pm 16\pm 3$&$ 780\pm 70\pm 30$&$0.283\pm0.014$\\
1.9625--1.9750&$ 209\pm 17\pm 3$&$ 850\pm 70\pm 32$&$0.288\pm0.013$\\
\end{tabular}
\end{ruledtabular}
\end{table*}
These form factor values are obtained as averages over mass-interval width.
The four measurements from PS170~\cite{LEAR} with lowest mass
are located within the first mass interval of Table~\ref{sumtab}.
Consequently, for the mass region near threshold,
where the results from PS170 indicate that the form factor changes rapidly 
with mass, we calculate the cross section
and effective form factor using a smaller mass-interval size.
These results are listed in Table~\ref{tablow}, and shown in   Fig.~\ref{fig19}.
From Figs.~\ref{fig17}, \ref{fig18}, and \ref{fig19},
it is evident that the \babar\ effective form factor results are in 
reasonable agreement with, and in general more precise than, those from previous 
experiments. However, in the region 1.88--2.15 GeV/$c^2$, the \babar\
results are systematically above those from the other experiments.

The form factor has a complex mass dependence.
The significant increase in the form factor as the $p\bar{p}$ threshold 
is approached may be due to final-state interaction between the proton and
antiproton~\cite{fsi1,fsi2,fsi3,fsi4}.
The rapid decreases of the form factor and cross section near 
2.2~GeV/$c^2$, 2.55~GeV/$c^2$, and 3~GeV/$c^2$
have not been discussed in the literature. The form-factor mass dependence
below 3~GeV/$c^2$ is not described satisfactorily by existing models
(see, for example, Refs.~\cite{ffmod1,ffmod2,ffmod3,ffmod4}).
The dashed curve in Fig.~\ref{fig18} corresponds to a fit of the asymptotic
QCD dependence of the proton form factor~\cite{QCD},  
$F_{p\bar{p}}\sim\alpha_s^2(M_{p\bar{p}}^2)/M_{p\bar{p}}^4\sim D/(M_{p\bar{p}}^4\log^2 (M_{p\bar{p}}^2/\Lambda^2))$,
to the existing data  with $M_{p\bar{p}}>3~{\rm GeV}/c^2$. Here 
$\Lambda=0.3$~GeV and $D$ is a free fit parameter. All the data above 
3~GeV/$c^2$ except the two points from Ref.~\cite{CLEO2012} marked ``NU'' 
are well described by this function. Adding the points
from Ref.~\cite{CLEO2012} changes the fit $\chi^2/\nu$ from 9/16 to 41/18,
where $\nu$ is the number of degrees of freedom. The 
measurement of Ref.~\cite{CLEO2012} indicates that 
the form factor at $M_{p\bar{p}}\approx 4~{\rm GeV}/c^2$ decreases more 
slowly than predicted by QCD.

\section{\boldmath The $J/\psi$ and $\psi (2S)$  decays to 
$p\bar{p}$ }\label{jpsi}
The  differential cross section for ISR production
of a narrow resonance (vector meson $V$),
such as $J/\psi$, decaying into the final state $f$ can be calculated
using~\cite{ivanch}
\begin{equation}
\frac{d\sigma(s,\theta_\gamma^\ast)}{d\cos{\theta_\gamma^\ast}}
=\frac{12\pi^2 \Gamma(V\to e^+e^-) {\cal B}(V\to f)}{m_V\, s}\,
W(s,x_V,\theta_\gamma^\ast),
\label{eqpsi}
\end{equation}
where $m_V$ and $\Gamma(V\to e^+e^-)$ are the mass and electronic
width of the vector meson $V$, $x_V = 1-{m_V^2}/{s}$,
and ${\cal B}(V\to f)$ is the branching fraction of $V$
into the final state $f$. Therefore, the measurement of the number of
$J/\psi \to p\bar{p}$ decays
in $e^+ e^- \to p\bar{p}\gamma$ determines the product of the
electronic width and the branching fraction:
$\Gamma(J/\psi \to e^+e^-){\cal B}(J/\psi \to p\bar{p})$.
\begin{figure*}
\includegraphics[width=.40\textwidth]{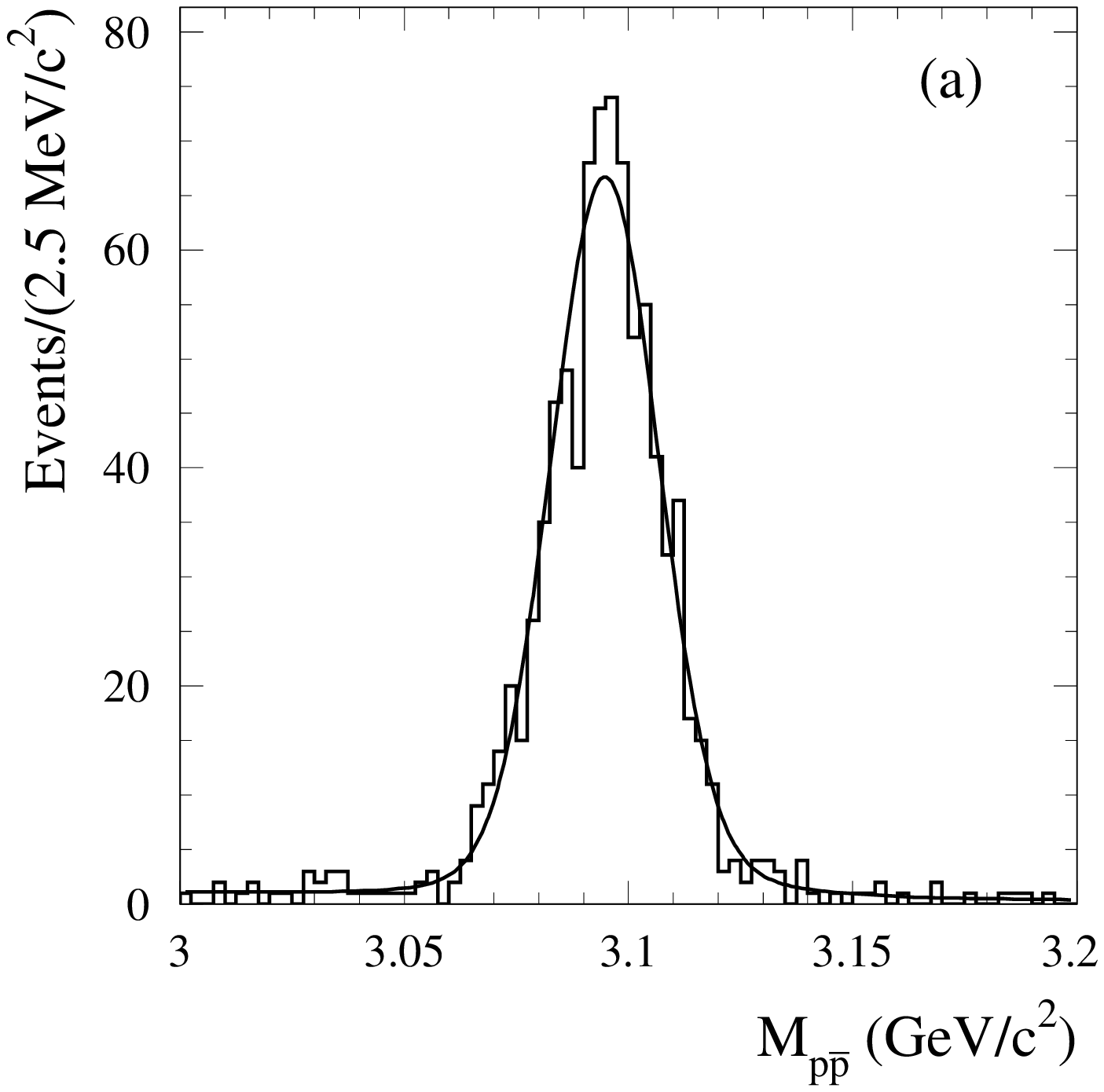}
\includegraphics[width=.40\textwidth]{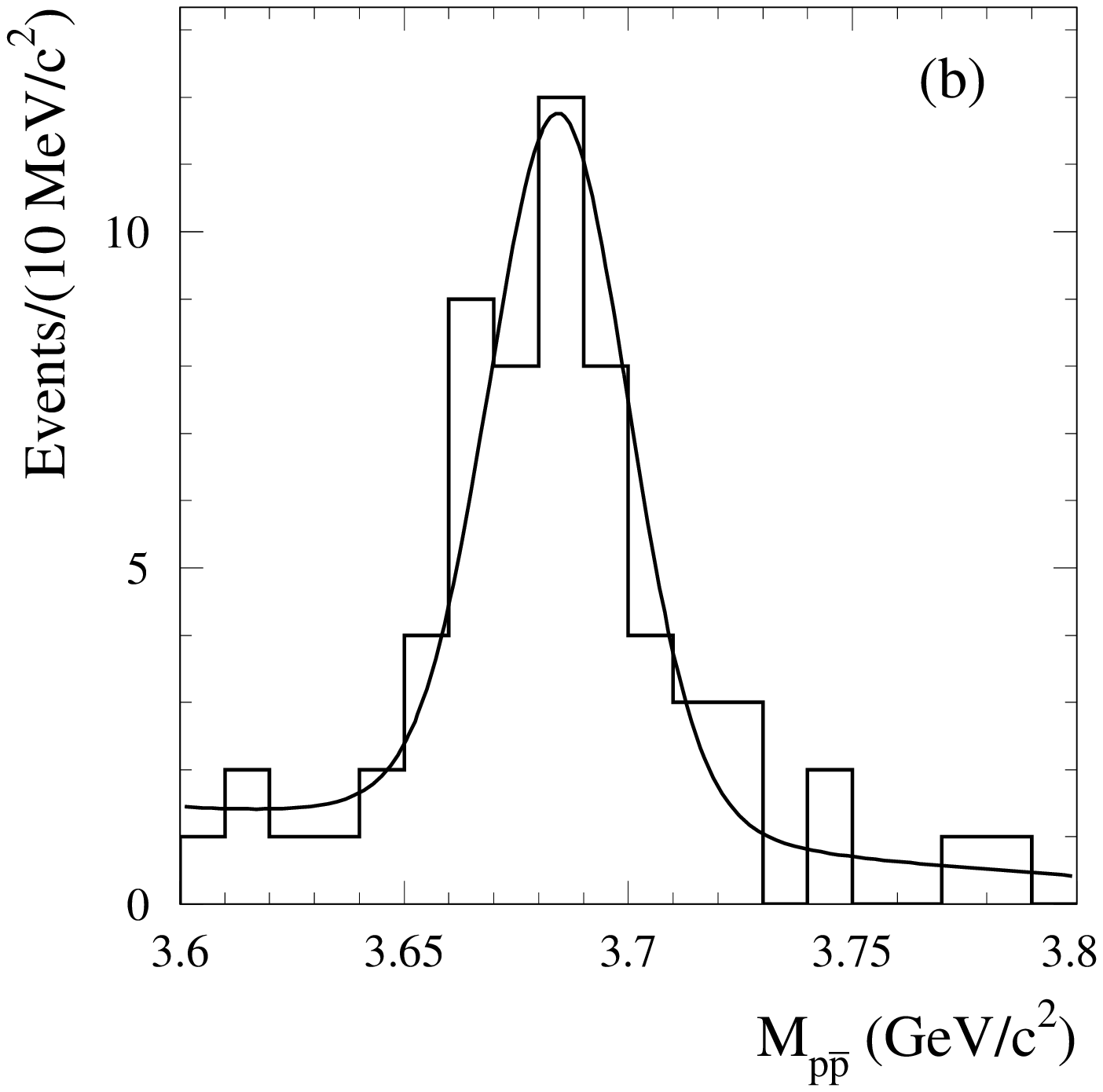}
\caption{The $p\bar{p}$ mass spectrum in the mass region
(a) near the $J/\psi$, and (b) near the $\psi(2S)$. The curves 
display the results of the fits described in the text.
\label{fig20}}
\end{figure*}
The $p\bar{p}$ mass spectra for selected events in the $J/\psi$ and
$\psi(2S)$ mass regions are shown in Fig.~\ref{fig20}.
To determine the number of resonance events, each spectrum
is fit with a sum of the probability density function (PDF)
for signal plus a linear background term.
The resonance PDF is
a  Breit-Wigner function convolved with
a double-Gaussian function describing detector resolution.
The Breit-Wigner widths and masses for the $J/\psi$ and
$\psi(2S)$ resonances are fixed at their nominal values~\cite{pdg}.
The parameters of the resolution function
are determined from simulation. 
To account for possible differences in detector response
between data and simulation,
the simulated resolution function is modified by
adding in quadrature  an additional $\sigma_G$ to both
the standard-deviation values of the double-Gaussian function, 
and introducing a shift of the central value of the resonance mass. 
The free parameters in the
fit to the $J/\psi$ mass region are the number of resonance events,
the total number of nonresonant background events, 
the slope of the background, $\sigma_G$, and the mass shift. 
In the $\psi(2S)$ fit the $\sigma_G$ and mass shift values
are fixed to those obtained for the  $J/\psi$.

The fit results are shown as the curves in Fig.~\ref{fig20}. 
We find $N_{J/\psi}=821\pm30$ and $N_{\psi(2S)}=43.5\pm7.7$;
the number of nonresonant events is $62\pm11$ for the 3--3.2~GeV/$c^2$
mass interval, and $19\pm6$ for the 3.6--3.8~GeV/$c^2$ interval.
These values are used to extract the nonresonant $e^+e^- \to p\bar{p}$
cross section. 
Since the background subtraction procedure for nonresonant
events (see Sec.~\ref{bkgsub}) uses events with $30<\chi^2_p<60$, 
the mass spectra obtained with this requirement must also be fit. The
numbers of $J/\psi$ and nonresonant events are found to be
$40\pm8$ and $19\pm7$. The ratio of $J/\psi$ events with
$30<\chi^2_p<60$ to the number with
$\chi^2_p<30$, $0.049\pm0.010$, is in good
agreement with the value of $\beta_{p\bar{p}\gamma}=0.049\pm0.001$ obtained
from simulation. In the $\psi(2S)$ mass region, no events are
found with $30<\chi^2_p<60$.  The remaining fit parameters are
$\sigma_G=5.0\pm1.0$ MeV/$c^2$ and
$M_{J/\psi}-M_{J/\psi}^{MC}=-(1.7\pm0.5)$~MeV/$c^2$.
The fitted value of $\sigma_G$ leads to an increase in simulation
resolution (11 MeV/$c^2$) of 10\%.

The corresponding detection efficiency values are estimated from MC simulation.
The event generator uses  
experimental information to describe the the angular
distribution of protons in $J/\psi$ and $\psi(2S)$ decay to $p\bar{p}$.
Specifically, each distribution is described by the dependence $1+a \cos^2\theta_p$,
with $a=0.672\pm0.034$ for $J/\psi$ decay~\cite{psipp1,BESpp}
and $a=0.72\pm 0.13$ for $\psi(2S)$ decay~\cite{E835psi,BESpsi2s}. 
The model error in the detection efficiency due to the uncertainty of
$a$ is negligible. The efficiencies are found to be
$\varepsilon_{\rm MC}$ = 0.174$\pm$0.001 for $J/\psi$ and
$\varepsilon_{\rm_MC}$ = 0.172$\pm$0.001 for $\psi(2S)$.
The fractional correction for the data-MC simulation differences discussed in 
Sec.~\ref{sdetef} is $-(3.6\pm 2.2)\%$.

The cross section for
$e^+e^-\to \psi\gamma\to p\bar{p}\gamma$ for
$20^\circ<\theta_\gamma^\ast<160^\circ$
is calculated as
\begin{equation}
\sigma(20^\circ<\theta_\gamma^\ast<160^\circ)=\frac{N_{\psi}}
{\varepsilon\, R\, L},
\end{equation}
yielding $(10.4\pm0.4\pm0.3)$ fb and $(0.55\pm0.10\pm0.02)$ fb
for the $J/\psi$ and $\psi(2S)$ states, respectively.
The radiative-correction factor $R=\sigma/\sigma_{Born}$ is
$1.007\pm0.010$ for the $J/\psi$ and $1.011\pm0.010$ for the $\psi(2S)$, where
both values are obtained from MC simulation at the generator level.

The total integrated luminosity for the data sample  is
$(469\pm 5)$ fb$^{-1}$.
From the measured cross sections and Eq.~(\ref{eqpsi}),
the following products are determined:
\begin{eqnarray}
{\Gamma(J/\psi\to e^+e^-)\,{\cal B}(J/\psi\to p\bar{p})=} \nonumber \\
(11.3\pm 0.4\pm 0.3)\mbox{ eV}, \nonumber \\
{\Gamma(\psi(2S)\to e^+e^-)\,{\cal B}(\psi(2S)\to p\bar{p})=}\nonumber \\
(0.67\pm0.12\pm0.02)\mbox{ eV}.
\end{eqnarray}
The systematic errors include the uncertainties on the detection efficiencies,
the integrated luminosity, and the radiative corrections.

Using the nominal values for the electronic widths~\cite{pdg},
we obtain the branching fractions
\begin{eqnarray}
{\cal B}(J/\psi\to p\bar{p})=(2.04\pm0.07\pm0.07)\times 10^{-3},\nonumber \\
{\cal B}(\psi(2S)\to p\bar{p})=(2.86\pm0.51\pm0.09)\times 10^{-4}.
\end{eqnarray}
These values are in agreement with the nominal values~\cite{pdg} of
$(2.17\pm0.07)\times 10^{-3}$ and $(2.76\pm0.12)\times 10^{-4}$, respectively,
and with the recent high-precision BESIII result~\cite{BESIII}
${\cal B}(J/\psi\to p\bar{p})=(2.112\pm0.004\pm0.031)\times 10^{-3}$.

\section{Summary}
The process $e^+e^-\to p\bar{p}\gamma$ has been studied for
$p\bar{p}$ invariant masses up to 4.5~GeV/$c^2$.
From the measured $p\bar{p}$ mass spectrum we extract the
$e^+e^-\to p\bar{p}$ cross section and the proton 
effective form factor.
The form factor has a  complex mass dependence.
The near-threshold enhancement of the form factor
observed in the PS170 experiment~\cite{LEAR} is confirmed in this study.
There are also three mass regions, near 2.2~GeV/$c^2$, 2.55~GeV/$c^2$, 
and 3~GeV/$c^2$, that exhibit steep decreases in the form factor 
and cross section. 

By analysing the proton angular distributions we  measure
the mass dependence of the ratio $|G_E/G_M|$ for $M_{p\bar{p}}$ 
from threshold to 3~GeV/$c^2$.
For masses up to 2.1~GeV/$c^2$, this ratio is found to be
significantly greater than unity,
in disagreement with the PS170 measurement~\cite{LEAR}.
The asymmetry in the proton angular distribution is found to be 
\begin{equation}
A_{\cos{\theta_p}}=-0.025\pm0.014\pm0.003 \nonumber
\end{equation}
for $M_{p\bar{p}}<3$~GeV/$c^2$.

From the measured event yields for
$e^+e^-\to J/\psi\gamma\to p\bar{p}\gamma$ and 
$e^+e^-\to\psi(2S)\gamma \to p\bar{p}\gamma$,
we determine the branching fraction values
\begin{eqnarray}
{\cal B}(J/\psi\to p\bar{p})=(2.04\pm0.07\pm0.07)\times 10^{-3}, \nonumber\\
{\cal B}(\psi(2S)\to p\bar{p})=(2.86\pm0.51\pm0.09)\times 10^{-4}. \nonumber
\end{eqnarray}

Our results on the cross section, form factors, and 
$J/\psi$ and $\psi(2S)$ decays agree with, and supersede, earlier \babar\ 
measurements~\cite{babarpp}.

\section{ \boldmath Acknowledgments}
We are grateful for the 
extraordinary contributions of our \pep2\ colleagues in
achieving the excellent luminosity and machine conditions
that have made this work possible.
The success of this project also relies critically on the 
expertise and dedication of the computing organizations that 
support \babar.
The collaborating institutions wish to thank 
SLAC for its support and the kind hospitality extended to them. 
This work is supported by the
US Department of Energy
and National Science Foundation, the
Natural Sciences and Engineering Research Council (Canada),
the Commissariat \`a l'Energie Atomique and
Institut National de Physique Nucl\'eaire et de Physique des Particules
(France), the
Bundesministerium f\"ur Bildung und Forschung and
Deutsche Forschungsgemeinschaft
(Germany), the
Istituto Nazionale di Fisica Nucleare (Italy),
the Foundation for Fundamental Research on Matter (The Netherlands),
the Research Council of Norway, the
Ministry of Education and Science of the Russian Federation, 
Ministerio de Ciencia e Innovaci\'on (Spain), and the
Science and Technology Facilities Council (United Kingdom).
Individuals have received support from 
the Marie-Curie IEF program (European Union) and 
the A. P. Sloan Foundation (USA).


\begin{thebibliography}{99}
\bibitem{BM} G.~Bonneau and F.~Martin, Nucl. Phys. B {\bf 27}, 381 (1971).
\bibitem{Coulomb} 
A.~B.~Arbuzov and T.~V.~Kopylova,
JHEP {\bf 1204}, 009 (2012).
\bibitem{kuhn_pp} H.~Czyz {\em et al.}, Eur. Phys. J. C {\bf 35}, 527 (2004).
\bibitem{DM1} B.~Delcourt {\em et al.} (DM1 Collaboration), Phys. Lett. B {\bf 86}, 395 (1979).
\bibitem{DM2} D.~Bisello {\em et al.} (DM2 Collaboration), Nucl. Phys. B {\bf 224}, 379 (1983);
Z. Phys. C {\bf 48}, 23 (1990).
\bibitem{FENICE} A.~Antonelli  {\em et al.} (FENICE Collaboration), Nucl. Phys. B {\bf 517}, 3 (1998).
\bibitem{ADONE73} M.~Castellano {\em et al.}, Nuovo Cim. A {\bf 14}, 1 (1973).
\bibitem{BES} M.~Ablikim {\em et al.} (BES Collaboration), 
Phys.\ Lett.\ B {\bf 630}, 14 (2005).
\bibitem{CLEO} T.~K.~Pedlar {\em et al.} (CLEO Collaboration), 
Phys.\ Rev.\ Lett.\  {\bf 95}, 261803 (2005).
\bibitem{CLEO2012} K.~K.~Seth, S.~Dobbs, Z.~Metreveli, A.~Tomaradze, T.~Xiao and G.~Bonvicini,
arXiv:1210.1596 [hep-ex].
\bibitem{LEAR} G.~Bardin  {\em et al.} (PS170 Collaboration), Nucl. Phys.  B {\bf 411}, 3 (1994).
\bibitem{E760} T.~A.~Armstrong {\em et al.} (E760 Collaboration), 
Phys. Rev. Lett. {\bf 70}, 1212 (1993).
\bibitem{E835} M.~Ambrogiani {\em et al.} (E835 Collaboration), Phys. Rev. D {\bf 60}, 032002
(1999); M.~Andreotti {\em et al.}, Phys. Lett. B {\bf 559}, 20 (2003).
\bibitem{babarpp} B.~Aubert {\em et al.} (\babar\ Collaboration),
Phys.\ Rev.\  D {\bf 73}, 012005 (2006).
\bibitem{ref:babar-nim} B.~Aubert {\em et al.} (\babar\ Collaboration),
Nucl. Instr. and Meth. A {\bf 479}, 1 (2002).
\bibitem{ifr} W.~Menges,
IEEE Nucl.\ Sci.\ Symp.\ Conf.\ Rec.\  {\bf 5}, 1470 (2006).
\bibitem{EVA} H.~Czy\.{z} and J.~H.~K\"uhn, 
Eur. Phys. J. C {\bf 18}, 497 (2001).
\bibitem{strfun} M.~Caffo, H.~Czy\.{z}, and E.~Remiddi,
Nuovo Cim. A {\bf 110}, 515 (1997);
Phys. Lett. B {\bf 327}, 369 (1994).
\bibitem{PHOTOS}
E.~Barberio and Z.~W\c{a}s, Comput. Phys. Commun. {\bf 79}, 291 (1994).
\bibitem{JETSET} T.~Sj\"ostrand, Comput. Phys. Commun. {\bf 82}, 74 (1994).
\bibitem{GEANT4} S.~Agostinelli {\em et al.},
Nucl. Instr. and Meth. A {\bf 506}, 250 (2003).
\bibitem{Chernyak} V.~L.~Chernyak, private communication.
\bibitem{TPEep}P.~G.~Blunden, W.~Melnitchouk and J.~A.~Tjion,
Phys. Rev. Lett. {\bf 91}, 142304 (2003).
\bibitem{Rosenbluth}J.~Arrington, Phys. Rev. C {\bf 69}, 022201 (2004).
\bibitem{polar1}O.~Gayou {\it et al.}, 
Phys. Rev. Lett. {\bf 88}, 092301 (2002).
\bibitem{polar2}V.~Punjabi {\it et al.}, Phys. Rev. C {\bf 71}, 055202 (2005);
erratum-ibid. C {\bf 71}, 069902 (2005).
\bibitem{polar3}A.~J.~R.~Puckett {\it et al.}, Phys. Rev. Lett. {\bf 104},
242301 (2010).
\bibitem{TPEee}E.~Tomasi-Gustafsson {\it et al.},
Phys. Lett. B {\bf 659}, 197 (2008).
\bibitem{fsi1}J.~Haidenbauer {\it et al.},
Phys.\ Lett.\ B {\bf 643}, 29 (2006).
\bibitem{fsi2}G.~Y.~Chen, H.~R.~Dong and J.~P.~Ma,
Phys.\ Lett.\ B {\bf 692}, 136 (2010).
\bibitem{fsi3}V.~F.~Dmitriev and A.~I.~Milstein,
Nucl.\ Phys.\ Proc.\ Suppl.\  {\bf 181-182}, 66 (2008).
\bibitem{fsi4}O.~D.~Dalkarov, P.~A.~Khakhulin and A.~Y.~Voronin,
Nucl.\ Phys.\ A {\bf 833}, 104 (2010).
\bibitem{ffmod1}M.~A.~Belushkin, H.-W.~Hammer and U.-G.~Meissner,
Phys.\ Rev.\ C {\bf 75}, 035202 (2007).
\bibitem{ffmod2}J.~P.~B.~C.~de Melo {\it et al.},
Phys.\ Lett.\ B {\bf 671}, 153 (2009).
\bibitem{ffmod3}S.~Furuichi, H.~Ishikawa and K.~Watanabe,
Phys.\ Rev.\ C {\bf 81}, 045209 (2010).
\bibitem{ffmod4}E.~L.~Lomon and S.~Pacetti,
Phys.\ Rev.\ D {\bf 85}, 113004 (2012);
erratum-ibid.\ D {\bf 86}, 039901 (2012).
\bibitem{QCD} V.~L.~Chernyak, A.~R.~Zhitnitsky, JETP Lett. {\bf 25}, 510 (1977);
G.~Lepage, S.~Brodsky, Phys. Rev. Lett. {\bf 43}, 545 (1979).
\bibitem{ivanch} M.~Benayoun {\em et al.}, Mod. Phys. Lett. A {\bf 14}, 2605 (1999).
\bibitem{pdg} J.~Beringer {\em et al.} (Particle Data Group), 
Phys. Rev. D {\bf 86}, 010001 (2012).
\bibitem{psipp1} D.~Pallin {\em et al.} (DM2 Collaboration), Nucl. Phys. B {\bf 292}, 653 (1987);
R.~Brandelik {\em et al.} (DASP Collaboration), Z. Phys. C {\bf 1}, 233 (1976);
I.~Peruzzi {\em et al.} (MARK I Collaboration), Phys. Rev. D {\bf 17}, 2901 (1978);
M.~W.~Eaton {\em et al.} (MARK II Collaboration), Phys. Rev. D {\bf 29}, 804 (1984).
\bibitem{BESpp}J.~Z.~Bai {\em et al.} (BES Collaboration), Phys. Lett. B {\bf 591}, 42 (2004).
\bibitem{E835psi} M.~Ambrogiani {\em et al.} (E835 Collaboration), 
Phys. Lett. B {\bf 610}, 177 (2005).
\bibitem{BESpsi2s} M.~Ablikim {\it et al.} (BES Collaboration),
Phys.\ Lett.\ B {\bf 648}, 149 (2007).
\bibitem{BESIII}M.~Ablikim {\it et al.}  (BESIII Collaboration),
Phys.\ Rev.\ D {\bf 86}, 032014 (2012).
\end{thebibliography}
\end{document}